\documentclass{aa}  

\usepackage{graphicx}
\usepackage{txfonts}
\usepackage{subcaption}
\usepackage{enumitem}
\usepackage{xcolor}
\usepackage{pdfpages}
\usepackage[flushleft]{threeparttable}
\usepackage{tablefootnote}

\usepackage{natbib,twoopt}
\bibpunct{(}{)}{;}{a}{}{,}             
\makeatletter
  \newcommandtwoopt{\citeads}[3][][]{\href{http://adsabs.harvard.edu/abs/#3}%
    {\def\hyper@linkstart##1##2{}%
     \let\hyper@linkend\@empty\citealp[#1][#2]{#3}}}
  \newcommandtwoopt{\citepads}[3][][]{\href{http://adsabs.harvard.edu/abs/#3}%
    {\def\hyper@linkstart##1##2{}%
     \let\hyper@linkend\@empty\citep[#1][#2]{#3}}}
  \newcommandtwoopt{\citetads}[3][][]{\href{http://adsabs.harvard.edu/abs/#3}%
    {\def\hyper@linkstart##1##2{}%
     \let\hyper@linkend\@empty\citet[#1][#2]{#3}}}
  \newcommandtwoopt{\citeyearads}[3][][]%
    {\href{http://adsabs.harvard.edu/abs/#3}
    {\def\hyper@linkstart##1##2{}%
     \let\hyper@linkend\@empty\citeyear[#1][#2]{#3}}}
\makeatother

\begin{document} 

\title{Quantitative polarimetry for the transition disk \\ 
in RX~J1604.3-213010
\thanks{The reduced images (FITS files) are available at the CDS via anonymous ftp to cdsarc.u-strasbg.fr (130.79.128.5) or via http://cdsarc.u-strasbg.fr/viz-bin/qcat?J/A+A/598/A43.}}

\author{J. Ma
          \inst{1}
          \and
          H.M. Schmid \inst{1}\fnmsep
          \and
          C. Tschudi \inst{1}\fnmsep
          }

   \institute{Institute for Particle Physics and Astrophysics, 
          ETH Zurich, Wolfgang Pauli Strasse 17, CH-8093 Zurich\\
              \email{jma@phys.ethz.ch}}

   \date{Received 16 December 2022 / Accepted 3 June 2023}

 
  \abstract
   {The characterization of the dust in protoplanetary disks is important for a better understanding of the resulting composition of forming planets and the dust particle evolution in these systems.}
   {We aim to accurately characterize the properties of the dust in the face-on transition disk around RX~J1604.3–213010 (RX~J1604) by analyzing the multiwavelength scattered light intensity and polarization images obtained with the ZIMPOL and IRDIS sub-instruments of VLT/SPHERE.}
   {We used archival data of RX~J1604 from the ESO archive and carefully corrected the polarization signal for instrumental effects, also taking the interstellar polarization into account. We measured the radial profiles of the disk for the azimuthal polarization, $Q_\varphi(r)$, in the R, J, and H bands and describe variations in our data due to the seeing and other effects. We derived the intrinsic polarization profiles of the disk, $\hat{Q}_\varphi(r)$, by comparing the data with rotationally symmetric models convolved with the point spread functions of the observations. We also measured the disk intensity, $I_{\rm disk}(r)$, with reference star differential imaging for the J and H bands. This provides the disk-integrated polarized intensity, $\hat{Q}_\varphi/I_\star$, for the R, J, and H bands and the averaged fractional polarization, $\langle \hat{p}_\varphi \rangle$, for the J and H bands. We investigated the azimuthal dependence of the scattered light and the shadows produced by hot dust near the star. The derived results were finally compared with model calculations to constrain the scattering properties of the reflecting dust in RX~J1604.}
  {RX~J1604 is a dipper source, and the data show different kinds of variability. However, a detailed analysis of repeated measurements shows that the results are not affected by dipping events or atmospheric seeing variations. 
  We derive accurate radial disk profiles for the intrinsic polarized intensity, $\hat{Q}_\varphi(r)/I_\star$, and measure different profile peak radii for different bands because of the wavelength dependence of the dust opacity. 
  The disk-integrated polarization is $\hat{Q}_\varphi/I_{\star}= 0.92\pm0.04\%$ for the R band and $1.51\pm0.11\%$ for the J band, indicating a red color for the polarized reflectivity of the disk. 
  The intensity of the disk is $I_{\rm disk}/I_\star= 3.9\pm 0.5~\%$ in the J band, and the fractional polarization is $\langle \hat{p}_\varphi \rangle =38\pm 4\%$ for the J band and $42\pm 2\%$ for the H band. 
  The comparison with the IR excess for RX~J1604 yields an apparent disk albedo of about $\Lambda_I \approx 0.16\pm 0.08$. We also find that previously described shadows seen in the R band data are likely affected by calibration errors. We derive, using dust scattering models for transition disks, approximate J band values for the scattering albedo $\omega\approx 0.5$, scattering asymmetry $g\approx 0.5$, and scattering polarization $p_{\rm max}\approx 0.7$ for the dust.}
   {The bright disk of RX~J1604 has a very simple axisymmetric structure and is therefore well suited as a benchmark object for accurate photo-polarimetric measurements. 
   We derive values for the disk polarization, $\langle \hat{p}_\varphi\rangle,$ and the apparent disk albedo, $\Lambda_I$, for the J band. Because $\langle \hat{p}_\varphi\rangle$ and $\Lambda_I$ depend predominantly on dust scattering parameters and only weakly on the disk geometry, 
   these parameters define tight relations for the dust scattering parameters between $\omega$ and $p_{\rm max}$ and between $\omega$ and $g$.
   The positive R to J band color for the polarized reflectivity, $(\hat{Q}_\varphi/I_\star)_{\rm J}\approx 1.64 \cdot (\hat{Q}_\varphi/I_\star)_{\rm R}$, is mainly a result of the wavelength dependence of dust parameters because the scattering geometry is expected to be very similar for different colors. 
   This work demonstrates the potential of accurate photo-polarimetric measurements of the circumstellar disk RX~J1604 for the determination of dust scattering parameters that strongly constrain the physical properties of the dust.}

   \maketitle
%
\section{Introduction}
Protoplanetary disks consisting of gas and dust are where planetary systems are born. The dust properties and evolution in these disks are closely related to planet formation. Protoplanetary disks were first detected through their strong IR excess in the spectral energy distribution (SED) of pre-main-sequence stars because the dust in the disk absorbs the stellar light and emits this energy as thermal emission \citep{Strom1989}. 
Observations of the scattered light from the disk were achieved with ground-based telescopes \citep[e.g.,][]{Fukagawa2006, Hashimoto2012, Muto2012, Quanz2013, Garufi2013, Benisty2015, Ginski2016, Avenhaus2018, Boccaletti2020} and with the \textit{Hubble} Space Telescope \citep[HST; e.g.,][]{Silber2000, Perrin09}. 
This method probes the small dust particles in the surface layers of the disk regions strongly irradiated by the central star. A popular technique for the investigation of the disk is polarimetric differential imaging because it is well suited to disentangle the polarized scattered light of the disk from the direct and therefore unpolarized light of the bright central star.
This technique has been implemented in several adaptive optics (AO) systems at large ground-based telescopes, such as Nasmyth Adaptive Optics System – Near-Infrared Imager and Spectrograph (NACO) at Very Large Telescope (VLT), Coronagraphic Imager with Adaptive Optics (CIAO) and High-Contrast Coronographic Imager for Adaptive Optics (HiCIAO) at Subaru Telescope, Gemini Planet Imager (GPI) at Gemini South Telescope, and Spectro-Polarimetric High-contrast Exoplanet REsearch (SPHERE) at VLT for high resolution and high contrast observations of circumstellar disks \citep[e.g.,][]{Schmid2022}. Many disks -- with a large diversity of disk morphologies and structures, such as gaps, rings, and spirals -- have been detected \citep[][and references therein]{Benisty22}. 

Transition disks are objects with a large inner cavity where the disk material has been depleted \citep[e.g.,][]{Espaillat2014}. This subgroup of disks is ideal for photo-polarimetric measurements because the outer disks surrounding the cavity are well resolved and many of these objects have been studied with a focus on characterizing and understanding the observed disk geometry. However, many fewer studies have been carried out on the characterization of the dust properties based on quantitative measurements of the scattered light, such as the disk intensity and fractional polarization. Accurate photo-polarimetric measurements only exist for a few disk systems, such as the bright extended disk AB Aur \citep{Perrin09} or the transition disks HD~34700, HD142527, and HD169142 \citep{Monnier2019, Hunziker2021, Tschudi21}. 
In addition, the relative wavelength dependence of the reflected intensity or polarized intensity has been derived in several systems, including TW Hya \citep{Debes2013}, HD~100546 \citep{Mulders2013}, and HD~135344B \citep{Stolker2016}. These disks typically show a higher reflectivity at longer wavelengths (with the exception of TW Hya) and a high degree of polarization ($p_{\rm disk}>20~\%$), possibly indicating the  presence of micron-sized dust aggregates, in agreement with the theory of dust coagulation. More such measurements for the dust in disks need to be collected for a better understanding of the dust evolution and dynamics in circumstellar disks as well as the planet formation process. 

In this work we present and analyze archival data of the transitional disk around RX~J1604.3–213010 (2MASS J16042165-2130284, hereafter RX~J1604) taken with the VLT instrument SPHERE \citep{Beuzit2019} for visual and near-IR wavelengths, with the main goal of providing accurate photometric and polarimetric measurements of the scattered light from this disk in order to characterize the scattering dust. 
In Sect. \ref{section:the-disk} we summarize observations and properties of this system from previous works. In Sect. \ref{section:data-desciption} we describe the used archival observations and the data reduction; details on the
polarimetric calibration are given in the appendix. Section \ref{section:analysis} presents our data analysis for the derivation of the intrinsic radiation signal of the disk. This includes a detailed fitting of observed radial disk profiles, $Q_{\varphi}(r)$, with point-spread-function-convolved disk models and the extraction of the disk intensity profile, $I_{\rm disk}(r),$ using reference star profiles. The initial polarimetric calibration of the data and the associated uncertainties are described in the appendix. As a result, we obtain disk-integrated values and radial and azimuthal profiles of the polarized flux, the disk-integrated scattered intensity, and the averaged fractional polarization. These results are combined with the literature value for the IR excess caused by the thermal emission of the dust. In Sect. \ref{section:discussion} these observational results are compared with model calculations for a transition disk from \citet{Ma2022} in order to constrain the dust scattering properties for the disk in RX~J1604. We discuss the value of quantitative measurements of the scattered light and the constraints on the dust parameters for the disk in RX J1604 in Sect.~\ref{sect:discussions}, and present our final conclusions in Sect.~\ref{sect:conclusions}.

\section{The RX~J1604.3–213010 disk}
\label{section:the-disk}
\begin{table*}[]
    \centering
    \caption{
    VLT/SPHERE polarimetric imaging data and corresponding observing conditions.}
    \begin{tabular}{l c c c c c c }
    \hline
    \hline
    Epochs & Object & nDIT $\times$ DIT & $n_{cycle}$  & $\tau_0$[ms] & seeing[$\arcsec$] & Air mass\\
    \hline
    \multicolumn{7}{l}{ZIMPOL R band}\\
    Jun 11, 2015 & RX~J1604 & 2 $\times$ 120 s & 6 & $2.4 \pm 0.1$ & $1.46 \pm 0.17$ & $1.01 \pm 0.01$\\
    \multicolumn{7}{l}{IRDIS H band}\\
    Jul 1, 2016\tablefootmark{a} & RX~J1604 & 2 $\times$ 64 s & 7 & $5.0\pm0.7$ & $0.58 \pm 0.09$ & $1.04 \pm 0.02$\\
    Jun 22, 2018& Hen3-1258 & 1 $\times$ 96 s & 4 & $7.1 \pm 1.0$ & $0.72 \pm 0.05$ & $1.07 \pm 0.01$\\
    \multicolumn{7}{l}{IRDIS J band}\\
    Aug 14, 2017 & RX~J1604 & 2 $\times$ 64 s      & 5 & $5.7 \pm 1.0$ & $0.61 \pm 0.08$ & $1.17 \pm 0.04$\\
     Aug 15, 2017 & RX~J1604 & 2 $\times$ 64 s      & 4 & $3.7 \pm 0.4$ & $0.91 \pm 0.13$ & $1.22 \pm 0.04$\\
     Aug 18, 2017 & RX~J1604 & 2 $\times$ 64 s      & 5 & $4.6 \pm 1.1$ & $0.53 \pm 0.06$ & $1.24 \pm 0.05$\\
     Aug 19, 2017 & RX~J1604 & 2 $\times$ 64 s      & 4 & $4.0 \pm 0.9$ & $0.66 \pm 0.07$ & $1.02 \pm 0.01$ \\
     Aug 22, 2017 & RX~J1604 & 2 $\times$ 64 s      & 7\tablefootmark{b} & $5.9 \pm 0.7$ & $0.56 \pm 0.10$ & $1.12 \pm 0.06$\\
     May 26, 2016 & HD~150193A & 6 $\times$ 16 s & 11 & $5.5 \pm 1.0$ & $0.45 \pm 0.07$ & $1.04 \pm 0.03$ \smallskip\\
     \multicolumn{7}{l}{not selected J band data}\\
     Sep 5, 2017 & RX~J1604 & 2 $\times$ 64 s      & 4 & $1.9 \pm 0.3$ & $0.98 \pm 0.16$ & $1.20 \pm 0.03$\\
     Sep 7, 2017\tablefootmark{a} & RX~J1604 & 2 $\times$ 64 s      & 4 & $3.6 \pm 0.3$ & $0.53 \pm 0.03$ & $1.35 \pm 0.05$\\
     Sep 17, 2017 & RX~J1604 & 2 $\times$ 64 s      & 4 & $2.6 \pm 0.4$ & $1.05 \pm 0.18$ & $1.37 \pm 0.05$\\
    \hline
    \end{tabular}
    \tablefoot{\tablefoottext{a}{
    Cloudy weather conditions according to the Astronomical site monitor (ASM)}\tablefoottext{b}{One cycle is discarded because the centering failed.}}
    \label{tab:obs-info}
\end{table*}

RX~J1604 is a pre-main-sequence star in the Upper Scorpius association with an apparent brightness of ${\rm G}= 11.7^m$ and ${\rm H}= 9.1^m$ \citep{Gaia2020, Cutri2003}, spectral type K2 and an estimated temperature of $T=4550~$K, and luminosity of $L=0.76~{\rm L}_\odot$ \citep{Preibisch99}. It is located at a distance of $145.7 \pm 0.4$ pc \citep{Gaia2020} and it has an age of 11$\pm$3~Myr as a member of the upper Sco \citep{Pecaut12}. The system shows a strong far-IR excess, a strong mid-IR minimum typical for transition disks, no silicate emission features around 10~$\mu$m, and flux variations in the near to mid IR range \citep{Dahm09}. 

Photometry in the visual revealed that RX~J1604 is an AA Tau variable or dipper star with aperiodic optical events consistent with the extinction by transiting dusty clouds near or in a compact, inner disk with $r \ll 1$~AU \citep[e.g.,][]{Ansdell2016}. High-cadence, multiwavelength light curves from the Rapid Eye Mount (REM) at La Silla showed a strong color dependence for a dimming event, with a strong attenuation in the g band of about 1~mag but only a weakening of 0.2 mag in the H band \citep{Sicilia2020}. These authors find that the observed wavelength dependences of the attenuation are similar, but not exactly equal, to the interstellar extinction law. 

Imaging at 880~$\mu m$ of the continuum emission \citep{Mathews2012, Zhang2014} and polarimetric imaging of the scattered radiation in the H band \citep{Mayama2012} revealed a bright, pole-on  disk ring with an inclination of $6^\circ$ and a radius of about $r\approx 0.43\arcsec$ (about 63~AU) with a large inner cavity. 
The polarized intensity shows in the data of \citet{Mayama2012} one surface brightness minimum or shadow in the ring at position angle (PA) $\sim 85^\circ$. A similar feature is also seen in the Zurich Imaging Polarimeter (ZIMPOL) R band data described by \citet{Pinilla2015} but at PA $\approx 46^\circ$, while later multi-epoch data obtained with The infrared dual-band imager and spectrograph (IRDIS) showed in the J band minima on the eastern and western side, which are variable in strength and position but always roughly located around $84^\circ$ and $266^\circ$ \citep{Pinilla2018}. 
They could explain the central star variability and the east and west minima on the outer disk by a highly inclined inner disk producing the dips in the stellar light curve and casting variable shadows on the almost pole-on outer transition disk.

\section{Data description}
\label{section:data-desciption}
We selected the polarimetric imaging data of RX~J1604 and corresponding calibrations taken with the VLT/SPHERE high contrast instrument \citep{Beuzit2019} using the ZIMPOL \citep{Schmid2018} and IRDIS 
\citep{Dohlen2008} focal plane imagers from the European Southern Observatory (ESO) archive for our analysis. Standard procedures for the data extraction and alignment, bias subtraction, flat-fielding, and bad pixel corrections were applied in the image reduction. The data were taken in units of polarimetric cycles, which include four Stokes exposures for linear polarization measurements, with half-wave plate positions $\theta=0^\circ$ and $45^\circ$ for Stokes $Q$ and $\theta=22.5^\circ$ and $67.5^\circ$ for Stokes $U$. For the polarimetric data, we also considered instrument and interstellar polarization, and the corresponding calibration is discussed in Appendix A.

As final products we obtained the differential Stokes $Q=I_{\rm 0}-I_{\rm 90}$ and $U=I_{\rm 45}-I_{\rm 135}$ images and the corresponding intensity images $I_Q=I_{\rm 0}+I_{\rm 90}$ and equivalent for $I_U = I_{\rm 45}+I_{\rm 135}$. In addition, we also calculate azimuthal Stokes parameters $Q_\varphi$ and $U_\varphi$ with respect to the central star as described in \citet{Schmid06} for radial Stokes parameters $Q_r$ and $U_r$ using the relation $Q_\varphi=-Q_r$ and $U_\varphi=-U_r$. The azimuthal polarization $Q_\varphi$ serves in our analysis as a proxy for the polarized flux $Q_\varphi\approx p\times I$ of the scattered radiation from the disk because $Q_\varphi$ is much less affected by a (strong) bias effect introduced by noisy data contrary to $p\times I$ \citep{Schmid06}.  

\subsection{ZIMPOL observations}
ZIMPOL data of RX~J1604 were obtained by \citet{Pinilla2015} on June 10, 2015, using the dichroic beam-splitter and the broadband R\_PRIM filters. The beam-splitter transmits the light from $\lambda=606$ to $698$~nm to ZIMPOL and reflects shorter and longer wavelengths to the wavefront sensor of the AO system. Therefore, the dichroic beam-splitter defines the pass-band of the polarimetric data $R_{\rm dic}$ with a central wavelength of $\lambda_c=652$~nm and $\Delta\lambda=92$~nm and not the wavelengths of the R\_PRIM filter as indicated by \citet{Pinilla2015}. 

Six complete polarimetric cycles were obtained with each Stokes exposure integrated by $2\times 120$~s, adding up to 96 minutes of total observing time. These data were taken in field stabilized mode, called the P2 mode, without coronagraph and without detector saturation. Therefore, they provide simultaneous stellar point spread function (PSF) observations (including a small, extended disk signal) and differential polarimetry for the scattered light from the disk.

The observations were taken under photometric conditions but the seeing was with an average value of $1.46\arcsec$ rather bad. In addition, the central star RX~J1604, which is used for the wavefront sensing for the AO system, is faint ($R\approx 11^m$) and a challenging target for visual observations with the SPHERE AO. Therefore, the normalized PSF can be characterized according to \cite{Schmid2018} by the parameters $ct_{n6}(0)/10^6 = 0.057\%, E_{f10}=4.6\%$, full width half maximum = 57.6mas, corresponding to an approximate Strehl ratios of only about 1.5~\%. 

After the basic data preprocessing, we corrected the instrumental polarization and find a residual polarization signal for the central star RX~J1604. As described in Appendix~\ref{App}, this polarization originates to a large extent from interstellar polarization, but potentially there is also a small contribution from an intrinsic polarization signal of the unresolved central object. 
It is important for the measurement of the scattered light from the circumstellar disk, to consider all these polarization components and estimate the possible impact of uncertainties on the derived disk polarization parameters. 

The R band images, the azimuthal polarization, $Q_{\varphi}(x,y),$ and the total intensity, $I_{\rm obs}(x,y),$ averaged for all six cycles from the night from June 11, 2015, are shown in Fig.~\ref{fig:qphi-itot-image}. The corresponding observational parameters are given in Table~\ref{tab:obs-info}.
\subsection{IRDIS observations}
Archival polarimetric imaging data for RX~J1604 with IRDIS are available for the J band ($\lambda=1245$ nm, $\Delta\lambda =240$ nm) and H band ($\lambda=1625$ nm, $\Delta\lambda =290$ nm). In addition, we also retrieved data from the objects Hen3-1258 and HD~150193A, which serve as PSF reference objects. Parameters for all the IRDIS data are summarized in Table~\ref{tab:obs-info} as well.

All the J and H band polarimetric observations with IRDIS were taken in field-stabilized mode with the N\_ALC\_YJH\_S coronagraph to avoid strong detector saturation. In addition, non-coronagraphic and unsaturated data were taken before and/or after the polarimetric observations for the flux and PSF calibration.

The J band data were taken over eight nights between August 22, 2017, and September 7, 2017, for a dedicated program to monitor the variable shadows on the disk ring \citep{Pinilla2018}. The same instrument configuration and procedures were used in all these nights, only the number of executed polarization cycles varied between four and seven. \\
We selected the data of the five best nights for our photo-polarimetric measurements. These data include for each night beside the long science integrations (DIT = 64~s), also short integration flux frames ($10 \times 0.8$~s) taken before, and longer integration flux frames ($10 \times 4$~s) with neutral density filter ND\_1.0 taken after all the polarimetric observation of one run. The excluded nights have lower quality because of less good observing conditions, variable total counts because of clouds or possible dimming events of the central star, and they all have only flux frames with ND\_1.0 filters, resulting in less precise photometric measurements.

Seven cycles of IRDIS-H polarimetry were obtained on July 1, 2016. These observations were preceded by short ($10\times 4$~s) non-coronagraphic flux calibration using the ND\_1.0 filter. We noticed flux variations during this run, which were most likely caused by thin clouds but a minor intrinsic dimming event for RX~J1604 cannot be excluded. 

The IRDIS polarimetry was reduced with the IRDAP pipeline software  \citep{vanHolstein2020}, which carries out the basic calibration steps (extraction, alignment, dark subtraction, flat fielding, bad pixel removal, etc.) and the polarimetric calibration based on a detailed Mueller matrix model of the telescope and the instrument. We find for the reduced RX~J1604 IRDIS data also a residual polarization with similar values as for the R band. This strongly indicates that there is an additional polarimetric signal from interstellar polarization or intrinsic polarization and we also correct for this (see Sect.~\ref{App}).

\begin{figure*}[]
    \centering
    \includegraphics[width=0.85\textwidth]{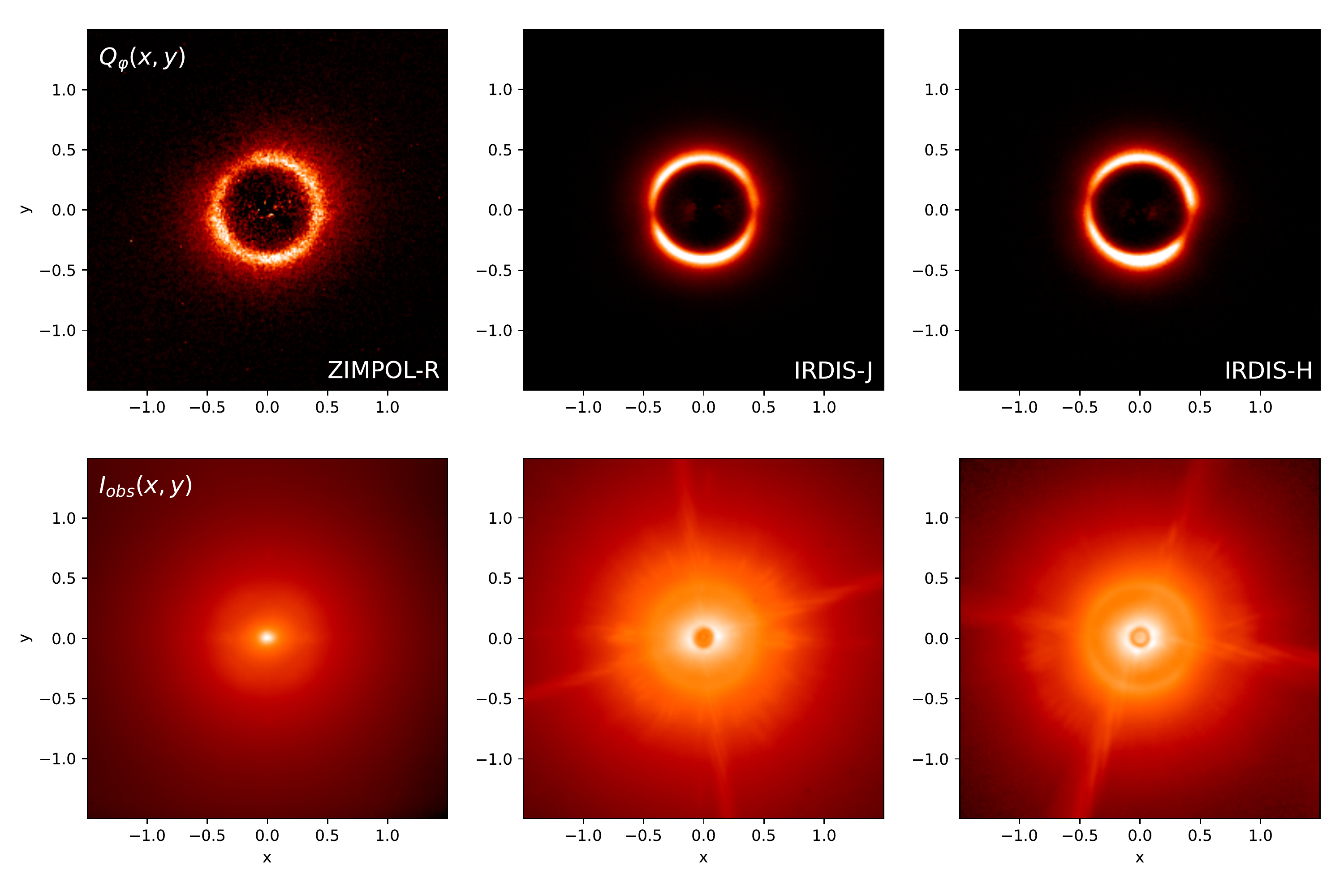}
    \caption{Averaged polarized intensity, $Q_{\varphi}(x,y)$ (upper), and total intensity, $I_{\rm obs}(x,y)$ (lower), for RX~J1604 taken with ZIMPOL in the R band (left), and with IRDIS in the J band (middle) and H band (right). $Q_\varphi$ is given on a linear color scale and $I_{\rm obs}$ on a logarithmic scale. The X and Y scales are in arcsec relative to the central star. North is up and east to the left.}
    \label{fig:qphi-itot-image}
\end{figure*}

Averaged polarization $Q_\varphi(x, y)$ images for the IRDIS-J band, including all the data from the five selected nights, and an averaged image for IRDIS-H band from July 1, 2016, are shown in Fig.~\ref{fig:qphi-itot-image}. The shadows on the east and west side of the disk ring are clearly visible in these $Q_\varphi(x,y)$ images. The intensity images $I_{\rm obs}(x,y)$ clearly show the size of the coronagraphic mask and one can also see the intensity signal of the disk ring, especially in the H band data.

\paragraph{Reference stars.}
To extract the disk intensity profile in our RX~J1604 data, we selected the polarimetric observations of Hen3-1258 for the H band and HD~150193A for the J band as reference stars from the ESO data archive. These observations were taken with the same IRDIS instrument setup and similar atmospheric conditions as our RX~J1604 data. Suitable PSF reference stars for coronagraphic polarimetry for the J and H bands are very rare and we found only useful data for targets that were observed as candidates for the presence of a circumstellar disk.

Deep polarimetric observations of HD~150193A are ideal because this target was suspected to have a disk by \cite{Meeus01} but deeper GPI data in the J band \citep{Monnier17} could not confirm the presence of a significant disk signal. Also, the selected reference star data from the ESO archive show no polarization signal from a disk. Thus, we use the 11 intensity frames resulting from the IRDAP reduction of HD~150193A as coronagraphic "PSF image" reference for our J band data of RX~J1604. Similarly for Hen3-1258 in the H band, which was observed with coronagraphic polarimetry, only very weak spiral features were found by \cite{Brown21} within 0.2$\arcsec$ from the central star. Therefore, we can use scaled intensity images of Hen3-1258 from the 4 polarimetric cycles like a reference star for the subtraction of the intensity of the central star in RX~J1604 at larger separations $> 0.2 \arcsec$. 

\section{Analysis}
\label{section:analysis}
\subsection{Integrated flux variability}
\label{section:integrated-intensity}
RX~J1604 is known as a dipper source or an AA Tau-type star and is suspected to have a strongly inclined inner disk, where transiting dust clouds produce flux minima on timescales of days \citep{Ansdell2016}.
Therefore, the central star is not a reliable flux reference for our data. The dips in RX~J1604 reduce the flux of the star by up to 1.0$^m$ or 1.5$^m$ in the V band, about 0.6$^m$ in the R band or 0.1$^m$ in the H band and the dimmings last between several hours and about a day \citep{Sicilia2020}. About a dozen such strong dips were found during 80 days of monitoring during the Kepler 2 mission \citep{Ansdell2016}. Thus, about 10~\% of the time, the RX~J1604 flux was significantly attenuated, especially at short wavelengths. 

We investigated the integrated stellar flux, $I_\star$, and the azimuthal polarization, $Q_\varphi$, of RX~J1604 for the presence of dimming events in our data. The detection of strong flux variations of $|\Delta {\rm m}/\Delta t|> 0.05$~mag/h in the R band, or $>0.02$~mag/h in the J and H bands or flux differences of $>0.1$~mag for the J band observations taken in different nights would point to a dimming event in our data. If no such variations are seen, then it seems safe to assume that the star brightness is close to the unattenuated brightness level, where ``close'' means about $\pm 10~\%$ for the R band and $\pm 5~\%$ for the J and H band \citep[see][]{Sicilia2020}.

We derive the photon counts with aperture photometry for individual measurements and calculate mean values $\langle I_\star \rangle$ and standard deviations $\sigma(I_\star)$ for a set of consecutive observations. 
We also measure the integrated azimuthal polarization $Q_\varphi$ of the disk because a dimming event of the central star by transiting dust should not change significantly the reflected light from the outer disk, while a variation in sky transparency would affect both the brightness of the star and the disk at the same level. As will be discussed below, the $Q_\varphi$ signal also varies systematically with the AO performance, which changes with the atmospheric seeing, but this effect can be recognized and corrected with the PSF calibrations.

\begin{table}[!ht]
    \centering
    \caption{Aperture count rates for the total intensity, $I_\star$, the azimuthal polarization, $Q_\varphi$, and the ratio, $Q_\varphi/I_\star$, for RX~J1604 for each observing date.}
    \label{tab:integrated-intensity-summary}
    \begin{tabular}{l c c c}
    \hline
    \hline
    \noalign{\smallskip}
    Epochs & 
    $\langle I_\star \rangle \pm \sigma(I_\star)$  &
    $\langle Q_\varphi \rangle \pm \sigma(Q_\varphi)$ &
    $Q_{\varphi}/I_{\star}$  \\
      & $\times 10^4$ s$^{-1}$ & $\times 10^4$ s$^{-1}$ & \% \\
    \noalign{\smallskip} 
    \hline
    \noalign{\smallskip\noindent ZIMPOL R band \hfill \smallskip}
    Jun 11, 2015 & 55.9 $\pm$ 0.5 & 0.246 $\pm$ 0.020 & 0.439 $\pm$ 0.033 \\
    \noalign{\smallskip\noindent IRDIS J band \hfill\smallskip}
    Aug 14, 2017 & 96.7\tablefootmark{a} & 1.043 $\pm$ 0.029 & 1.08 $\pm$ 0.03\\
    Aug 15, 2017 & 91.6\tablefootmark{a} & 1.051 $\pm$ 0.044 & 1.15 $\pm$ 0.05\\
    Aug 18, 2017 & 96.5\tablefootmark{a} & 1.122 $\pm$ 0.024 & 1.16 $\pm$ 0.03\\
    Aug 19, 2017 & 93.3\tablefootmark{a} & 1.117 $\pm$ 0.029 & 1.20 $\pm$ 0.03\\
    Aug 22, 2017 & 96.4\tablefootmark{a} & 1.157 $\pm$ 0.020 & 1.20 $\pm$ 0.02\\
    mean       & 94.9 $\pm$ 2.1 & 1.104 $\pm$ 0.054 & 1.16 $\pm$ 0.08   \\
    \noalign{\smallskip\noindent IRDIS H band
    \hfill\smallskip}   
    Jul 1, 2016 & 123. $\pm$ 18.\tablefootmark{b} & 1.40 $\pm$ 0.21 & (1.13 $\pm$ 0.34) \\
    \hline
    \end{tabular}
    \tablefoot{Average values and standard deviations are given for sets of identical measurements: 
        \tablefoottext{a}{single intensity measurement}, 
        \tablefoottext{b}{single measurement with an adopted relative uncertainty comparable to the variation $\sigma(Q_{\varphi})/\langle Q_{\varphi}\rangle$ of the same night.} }
\end{table}

The R band data are non-coronagraphic, and each of the six polarimetric cycles provides an $I_\star$ and a $Q_\varphi$ measurement. 
We derived for $I_\star$ the count rates of $I_{\rm obs}(x, y)$ in a circular aperture with $r=1.5\arcsec$. A very weak dark current and background level can be determined and subtracted assuming radial profiles fall off like a power law $I_\star(r) \propto r^{-\alpha}$ for the PSF seeing halo outside the AO control radius. The six cycles give a very small relative standard deviation $\sigma(I_\star)/\langle I_{\star}\rangle=0.9~\%$ (Table~\ref{tab:integrated-intensity-summary}) covering an observing sequence of 103 minutes. We conclude that the flux of RX~J1604 for this run is not attenuated by a dip. 

The polarized disk intensity $Q_\varphi$ is integrated in an annular aperture with $0.18\arcsec<r<1.5\arcsec$ and no background signal is visible in the differential polarization images. The annular aperture avoids the very noisy and low signal region at the location of the central star. The six cycles give $Q_\varphi$ signals with a relative standard deviation of $\sigma(Q_\varphi)/\langle Q_{\varphi}\rangle=8.3\%$ (Table~\ref{tab:integrated-intensity-summary}) ,
which can be explained by the variable polarimetric cancellation effects caused by 
the PSF smearing (see Sect.~\ref{section:radial-profile}).

Sensitive IRDIS polarimetry in the J band was taken in coronagraphic mode and therefore, these data lack simultaneous stellar intensity information. 
However, short non-coronagraphic flux and PSF calibration frames without saturation of the central star were taken before and after the polarimetric cycles. The flux frames after the polarimetry were obtained with a neutral density filter and therefore, the signal is $\sim$16 times weaker and because of detector noise not suited for accurate flux measurements. Nonetheless, these data confirm that no strong flux variations happened during an individual night. 
Interesting is the comparison of the count rates in a $r=1.5\arcsec$ aperture of the flux frames from the five selected nights in August 2017 taken before the polarimetric cycles, which yield a very small relative standard deviation of $\sigma(I_\star)/\langle I_{\star} \rangle \approx 2.2~\%$. This indicates that the selected J band observations were not significantly affected by dimming events. The $Q_\varphi$ signals measured in the annular aperture also show only small variations that are compatible with no significant variations, $\approx 5~\%$, in sky transparency or disk brightness. 

The IRDIS H band data from July 1, 2016, include only one flux calibration, and therefore we have no information about temporal variation from non-coronagraphic observations. However, the $Q_{\varphi}$ flux in the 7 polarimetric cycles varies strongly with a $\sigma({Q_\varphi})/\langle Q_{\varphi}\rangle = 15\%$ or by a factor of 1.7 between the highest and lowest value. This indicates that either an intrinsic dimming event or variable atmospheric transmission by clouds occurred, which is further discussed in Sect.~\ref{section:correlation}. 
Because we have no repeated flux measurement, we adopt in Table~\ref{tab:integrated-intensity-summary} the measured $\sigma({Q_\varphi})/\langle Q_{\varphi}\rangle = 15~\%$ deviation variability as systematic uncertainty for the H band $I_\star$-flux measurement $\sigma(I_\star)/I_\star$ and the sum of these two relative errors as uncertainty for the $Q_\varphi/I_\star$ parameter. 

\subsection{Radial profile analysis}
\label{section:radial-profile}
Accurate photo-polarimetry of circumstellar disks must take the variable PSF smearing and polarimetric cancellation effects of AO observations into account as pointed out and described in \citet{Tschudi21}. 
This is particularly important for observations with strongly variable PSFs, as is the case for our R band observations of RX~J1604. This problem is strongly reduced by the fact that the ZIMPOL observations simultaneously provide the PSF for the central star and the disk polarization signal. In addition, RX~J1604 is a pole-on disk with a circular ring and as a first approximation, we can carry out an analysis for radial profiles with much-enhanced S/N-ratios because of the azimuthal averaging \citep{Tschudi21}. However, the shadows cast by the inclined inner disk produce clear departures from rotational symmetry for the azimuthal flux distribution of the disk in RX~J1604, and therefore, we investigate the impact of this effect in Sect.~\ref{section:azimuthal_variation}.
 
\subsubsection{Model fits for the polarized intensity}
\label{section:polarized-intensity-model}
To account for the PSF smearing effects in our observations, we use a parametric model for the disk in RX~J1604, and search for the solution, which after PSF convolution fits best the observations. 
First, we derive the azimuthally averaged polarized intensity profiles $Q_\varphi(r)$ from the observations for the R, J, and H bands and adopt an intrinsic disk model $\Hat{Q}_{\varphi}(r)$ described by two power laws, 
\begin{equation}
    \centering
    \frac{\hat{Q}_{\varphi}(r)}{I_\star} = A_r\cdot\left(\left(\dfrac{r}{r_0}\right)^{-2\alpha_{in}}+\left(\dfrac{r}{r_0}\right)^{-2\alpha_{out}}\right)^{-\frac{1}{2}}
    \label{eq: two-power}
,\end{equation} 
with amplitude $A_r$, approximate peak radius $r_0$, and the two power-law indices $\alpha_{\rm in}>0$ and $\alpha_{\rm out}<0$ for the inner and outer disk slope, respectively.

The scattering polarization is assumed to be strictly in the azimuthal direction and hence $\Hat{U}_{\varphi}=0$. We convert the model profile $\Hat{Q}_{\varphi}(r)$ into a 2D image $\Hat{Q}_{\varphi}(x,y)$, then transform this into Stokes images $\Hat{Q}(x,y)$ and $\Hat{U}(x,y)$, which are convolved to $\Tilde{Q}(x,y)$ and $\Tilde{U}(x,y)$ using the appropriate PSF intensity images $I_\star(x,y)$.
For ZIMPOL, $I_\star(x,y)$ are the simultaneous intensity frames from the polarimetric cycles and for IRDIS we use the flux frames taken before the polarization measurements. 
The convolved Stokes images are transformed into an azimuthal polarization image $\Tilde{Q}_\varphi(x,y)$ from which the radial profile $\Tilde{Q}_\varphi(r)$ is calculated for the comparison with the observed profile $Q_\varphi(r)$. We then search for the model parameters that fit best the observations by minimizing the residuals of each radial point weighted by $r$ to account for the increasing disk area $2\pi r\,\Delta r$. 

\begin{figure}
    \centering
    \begin{subfigure}[]{0.45\textwidth}
        \includegraphics[width=\textwidth]{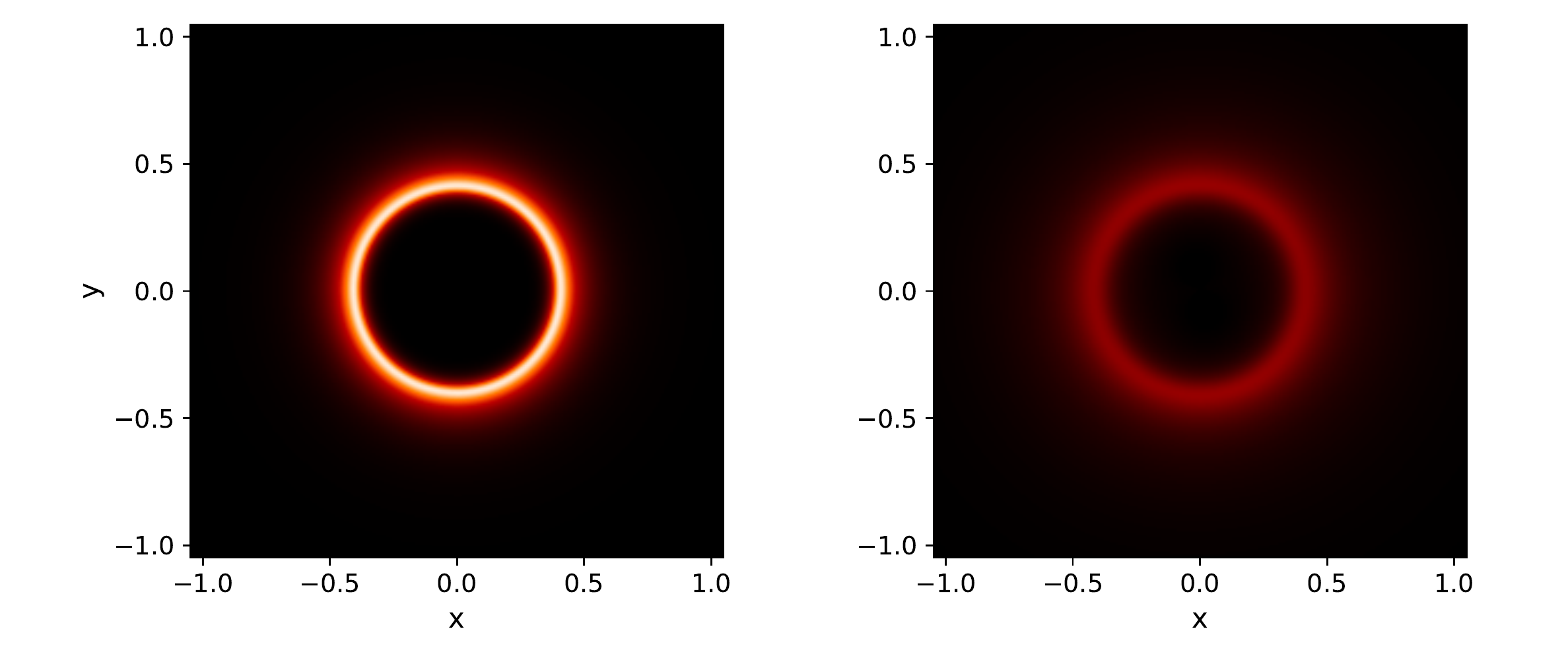}
    \end{subfigure}
    \caption{Disk model image for the derived intrinsic polarized intensity, $\hat{Q}_{\varphi}(x,y),$ for RX~J1604 in the R band (left) and the corresponding image, $\tilde{Q}_{\varphi}(x,y),$ after PSF convolution (right) with the observed intensity profile, $I_\star(x,y)$.}
    \label{fig:model-zimpol}
\end{figure}

\begin{figure}
    \centering
    \begin{subfigure}[]{0.9\textwidth}
        \includegraphics[width=0.48\textwidth]{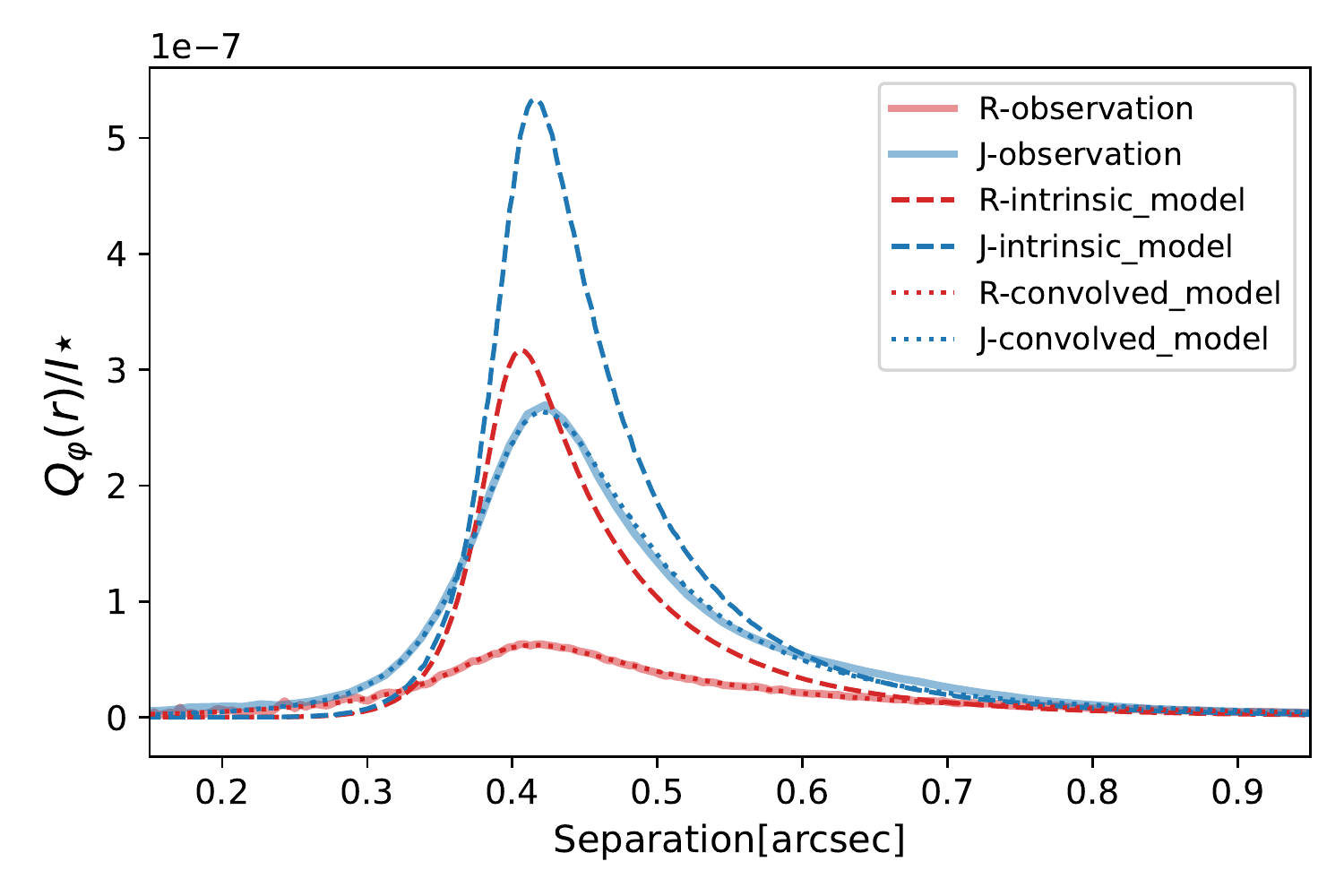}
    \end{subfigure}
    \caption{Comparison of the azimuthally averaged radial profiles of the polarized intensity for RX~J1604 derived for the observation, $Q_\varphi(r)/I_\star$, the intrinsic disk model, $\hat{Q}_\varphi(r)/I_\star$, and the convolved model, $\tilde{Q}_\varphi(r)/I_\star$, for the R and J bands. Units are given as relative flux per pixel of $3.6 {\rm mas} \times 3.6 {\rm mas}$. }
    \label{fig:rad-model}
\end{figure}

The main effects of the convolution are illustrated in Fig.~\ref{fig:model-zimpol} with a comparison of the intrinsic $\hat{Q}_\varphi(x,y)$ and convolved model $\tilde{Q}_\varphi(x,y)$ for the R band, and in Fig.~\ref{fig:rad-model} for the intrinsic $\hat{Q}_\varphi(r)$ and observed ${Q}_\varphi(r)$ profiles for the R band and J band. These examples show the reduction of the spatial resolution, the strong reduction of the peak surface brightness of up to a factor of 5 for the R band, and the enhanced polarization halo. The profiles for the convolved models $\tilde{Q}_\varphi(r)$ are practically indistinguishable from the observed profiles $Q_\varphi(r)$.

We note that for the PSF, we used the total intensity profile of the RX~J1604 system $I_{\rm PSF}=I_\star=I_{\rm star}+I_{\rm disk}$,  which is the flux of the central (unresolved) object $I_{\rm star}$ including the flux of the disk $I_{\rm disk}$. Thus, we overestimate the radial spread of $I_{\rm PSF}(r)$ by a few percent, particularly at the separation of the disk. This produces an over-correction of the PSF convolution and therefore, overestimates the polarized flux $\hat{Q}_{\varphi}$ for the intrinsic model. However, this effect is compensated for the ratio $\hat{Q}_{\varphi}/I_{\star}$ because we also overestimate the stellar flux by using $I_\star$ instead of $I_{\rm star}$. 
We estimated this effect with numerical simulations and find that the R band ratio $\hat{Q}_{\varphi}/I_{\star}(R)$ from Table~\ref{tab:fitting-parameters} derived with the simplified procedure differs by $1~\%$ from the more complicated treatment, where $I_{\rm disk}$ is first subtracted from $I_{\rm PSF}$ for the modeling of the convolution and from $I_\star$ for the evaluation of the ratio $\hat{Q}_{\varphi}/I_{\star}$. This difference is much less than other uncertainties, and therefore we simply use ratios of disk radiation parameters relative to the total system flux $I_\star$.

\paragraph{Resulting intrinsic disk polarization.} The best fitting parameters and 1$\sigma$ standard deviation for the intrinsic disk polarization profiles $\hat{Q}_{\varphi}(r)/I_{\star}$ are given in Table~\ref{tab:fitting-parameters} for the R, J, and H bands.The indicated uncertainties for the fitting parameters are based on the covariance matrix derived in the fitting procedure for the radial profile. 
This does not include flux calibration uncertainties for the parameter A, because the derived profiles are used to quantify PSF convolution effects and profile differences between different wavelengths The profiles can be used to derive the disk-integrated polarized intensities, $\hat{Q}_{\varphi}/I_\star$, which are also given in Table~\ref{tab:fitting-parameters}. However, these $\hat{Q}_{\varphi}/I_\star$ values include the flux calibration errors, which are important for the determination of these observational values. The uncertainties for $\hat{Q}_{\varphi}/I_\star$ are derived from the standard deviation from a set of individual measurements and calibrations.
The uncertainties are particularly large for the H band, because of non-photometric atmospheric conditions (see Sect.~\ref{sec:details}).

The intrinsic, disk-integrated $\hat{Q}_{\varphi}/I_\star$ values are for the R band roughly a factor of 2.1 higher, and for the J band about a factor of 1.3 higher than the observed values $Q_{\varphi}/I_\star$ given in Table~\ref{tab:integrated-intensity-summary} because of the PSF smearing and polarimetric cancellation. This stresses the importance of these effects for the derivation of the intrinsic polarization signal of circumstellar disks. 

There is a significant difference for $\hat{Q}_{\varphi}/I_\star$, of a factor of 1.64, between the J and R band, which defines a red color for the polarized reflectivity of the disk. We adopted a relative color gradient definition, $\eta$, from \cite{Tazaki2019} for the $\lambda$ interval between the selected R and J bands:\begin{equation}
    \eta_{R/J} = \dfrac{\log(\hat{Q}_{\varphi}/I_{\star})_{J}-\log(\hat{Q}_{\varphi}/I_{\star})_{R}}{\log(\lambda_J/\lambda_R)}\ = 0.76 \pm 0.18, 
    \label{Eq:refgradient}  
\end{equation}
where $\eta>0.5$ is classified as red in color.

There exist no previous polarized reflectivity determinations for the disk in RX~J1604. Compared to other transition disks the intrinsic R band value $\hat{Q}_{\varphi}/I_\star$ for RX~J1604 is about a factor of 2 higher than for HD~142527 \citep{Hunziker2021}, or HD~169142 \citep{Tschudi21}. In the $Q_{\varphi}/I_\star$ compilation for 75 protoplanetary disks of \citet{Garufi2022}, RX~J1604 would be in the top 20~\%. 
The red color for the reflectivity $\eta_{R/J}$ for RX~J1604 (Eq.~\ref{Eq:refgradient}) is very similar to the values $\eta_{VBB/H}=1.06\pm 0.14$ measured for HD~142527 \citep{Hunziker2021} and $\eta_{R'/I'}=1.05\pm 0.18$ for HD~169142 \citep{Tschudi21}. And also for HD~135344B \citep{Stolker2016}, HD~34700A \citep{Monnier2019}, and IM Lup \citep{Avenhaus2018}, a red wavelength dependence for $Q_{\varphi}/I_\star$ has been reported. In many cases, the derived colors are compatible with a gray reflectivity \citep{Avenhaus2018} because of the rather large uncertainties for the derived $\hat{Q}_{\varphi}/I_\star$ values. More measurements for the polarized reflectivity color $\eta$ are required to put the result for RX J1604 into context and such a study is in preparation \cite{Ma_prep}.

\begin{table}[]
    \centering
    \caption{RX~J1604 disk fitting parameters for the intrinsic polarized intensity profile, $\hat{Q}_\varphi/I_\star$, and derived  integrated disk radiation parameters (intrinsic) for the R, J, and H bands.}
    \begin{tabular}{c c c c}
    \hline\hline
                  & R & J & H \\
    \hline
    \multicolumn{4}{c}{Azimuthally averaged profiles $\hat{Q}_\varphi(r)/I_{\rm \star}$ 
    }
    \\
    A [$10^{-7}$] \tablefootmark{a} & 4.28 $\pm$ 0.10   & 7.23 $\pm$ 0.17  & 8.31 $\pm$ 0.28\\
    $r_0$ [AU]                & 59.7 $\pm$ 0.4    & 61.3 $\pm$ 0.3   & 62.0$\pm$ 0.4 \\
    $\alpha_{in}$             & 15.4 $\pm$ 1.6    & 14.9 $\pm$ 1.1   & 18.7 $\pm$ 2.5 \\
    $\alpha_{out}$            & -6.20 $\pm$ 0.05  & -6.72$\pm$ 0.07  & -7.43 $\pm$ 0.11\\
    \multicolumn{4}{c}{Disk integrated radiation parameters [\%]}\\
    $\Hat{Q}_{\varphi}/I_{\star}$ & 0.92 $\pm$ 0.04 & 1.51 $\pm$ 0.11 & 1.52 $\pm$ 0.45\\
    $I_{disk}/I_{\star}$\tablefootmark{b} & - & 3.9 $\pm$ 0.5 & 3.8 $\pm$ 0.8\\
    $\hat{p}_{\varphi}$\tablefootmark{b} & - & 38 $\pm$ 4 & 42 $\pm$ 2\\
    \hline
    \end{tabular}
    \tablefoot{\tablefoottext{a}{relative flux $\hat{Q}_\varphi(r)/I_\star$ 
    per pixel of 3.6~mas $\times$ 3.6~mas,} 
    \tablefoottext{b}{$I_{disk}/I_{\star}$ and $\hat{p}_{\varphi}$
    adopted from the derivation in Sect.~\ref{section:disk-intensity}.}}
    \label{tab:fitting-parameters}
\end{table}

\subsubsection{Details on the determination of the intrinsic profile}
\label{sec:details}
\begin{figure}
    \centering
    \begin{subfigure}[]{0.45\textwidth}
        \includegraphics[width=0.9\textwidth]{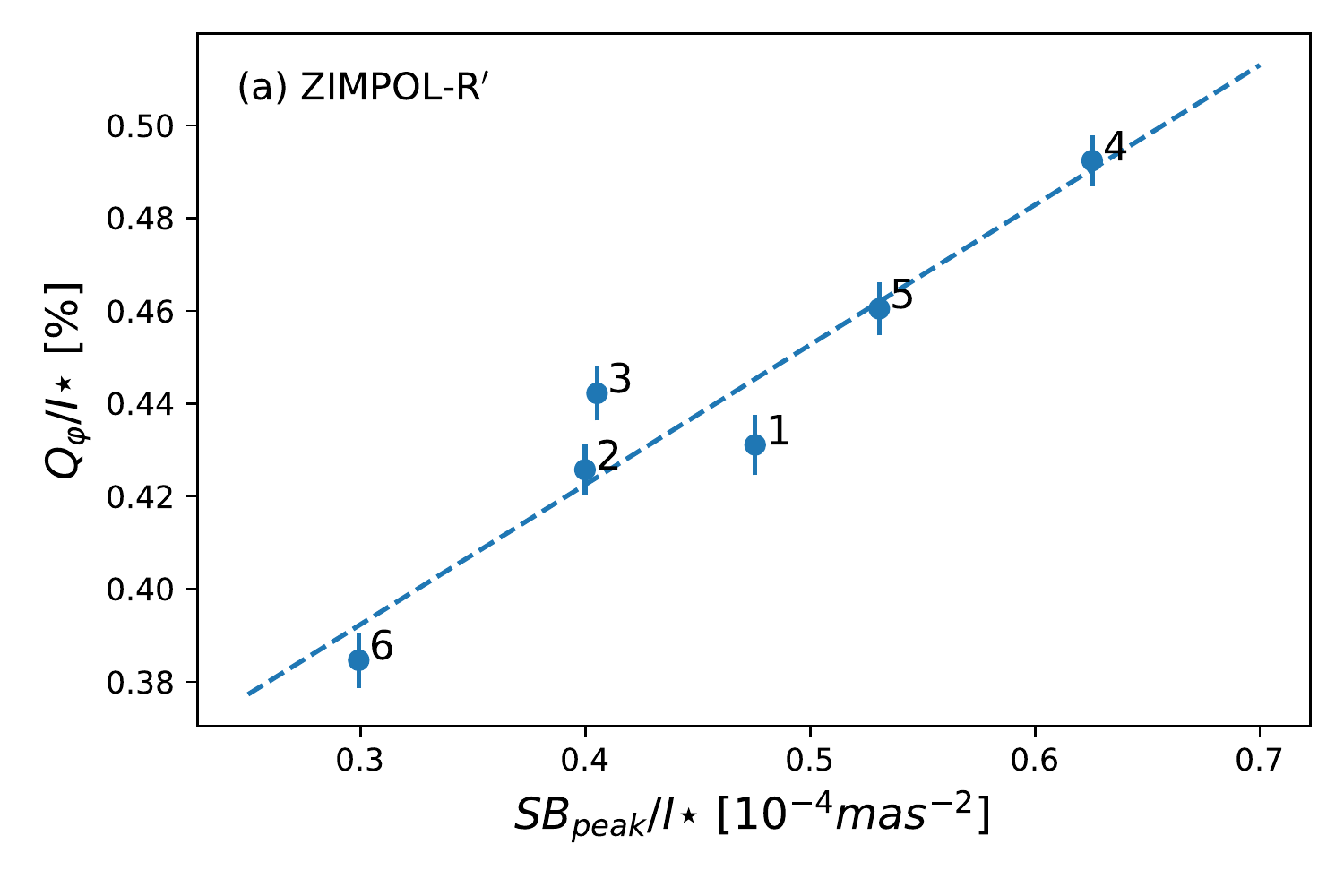}
        \vspace{-0.2cm}
        \label{fig:corr-qphi-peak-zimpol}
    \end{subfigure}
    \begin{subfigure}[]{0.45\textwidth}
        \includegraphics[width=0.9\textwidth]{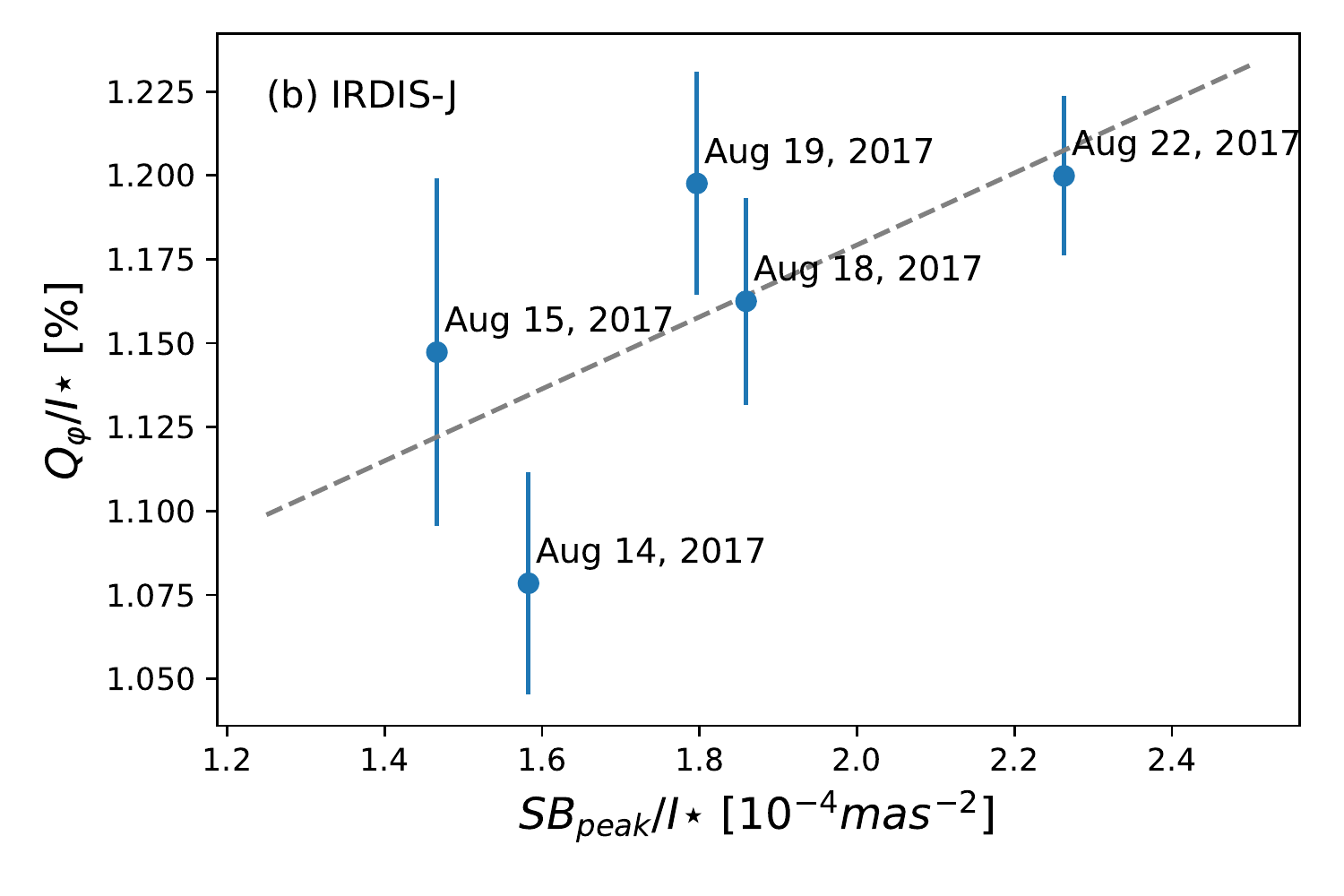}
        \vspace{-0.2cm}
        \label{fig:corr-qphi-peak-irdis}
    \end{subfigure}
    \caption{Dependence between the measured polarized disk flux, $Q_{\varphi}/I_{\star}$, and the stellar peak flux, ${\rm SB}_{peak}/I_{\star}$, for the six R band cycles from June 11, 2015 (top) and the five selected J band runs from August~2018 (bottom).} 
    \label{fig:corr-qphi-peak}
\end{figure}

\begin{figure}
    \centering
    \begin{subfigure}[]{0.45\textwidth}
        \includegraphics[width=\textwidth]{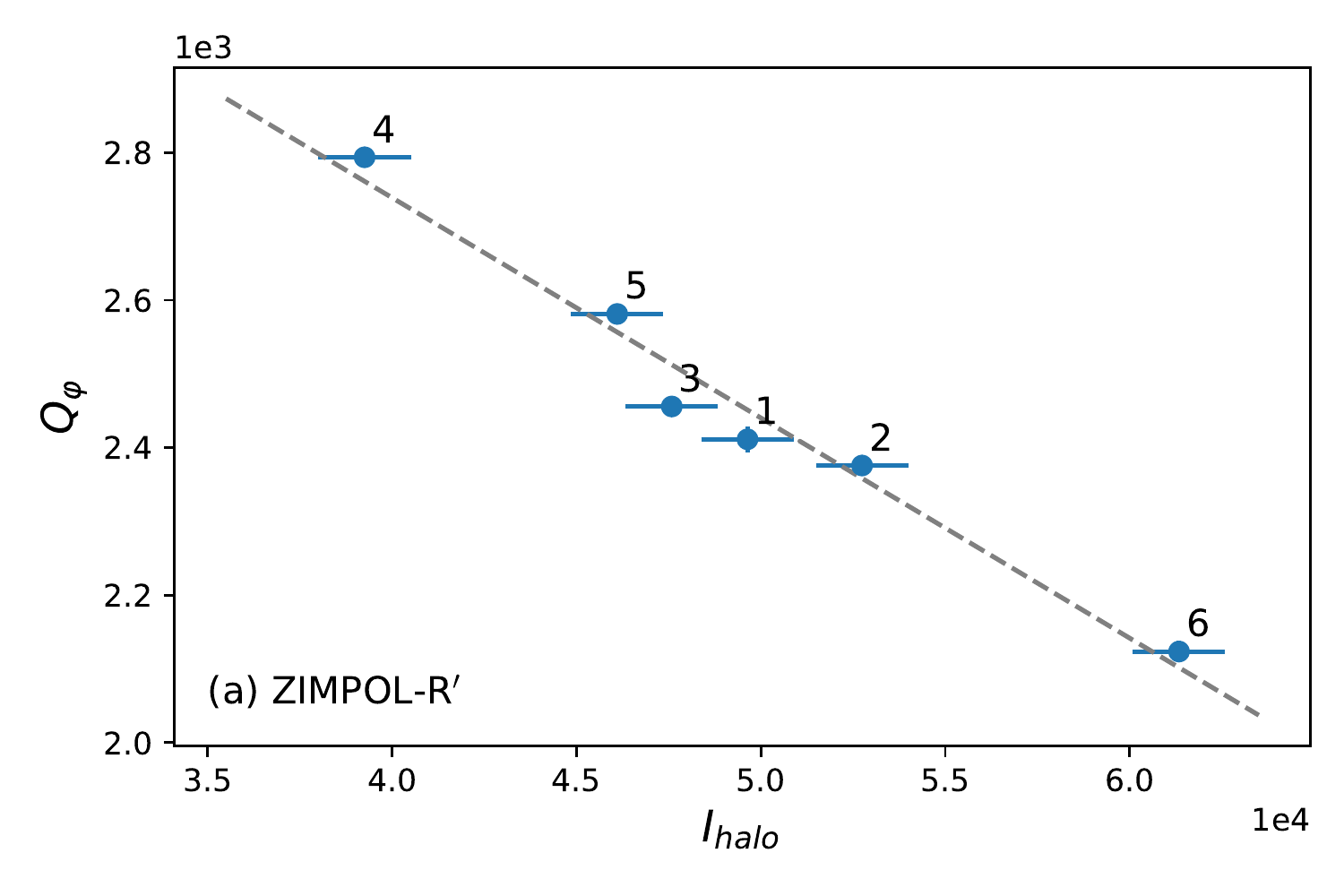}
        \label{fig:corr-qphi-halo-zimpol}
    \end{subfigure}
    \begin{subfigure}[]{0.45\textwidth}
        \includegraphics[width=\textwidth]{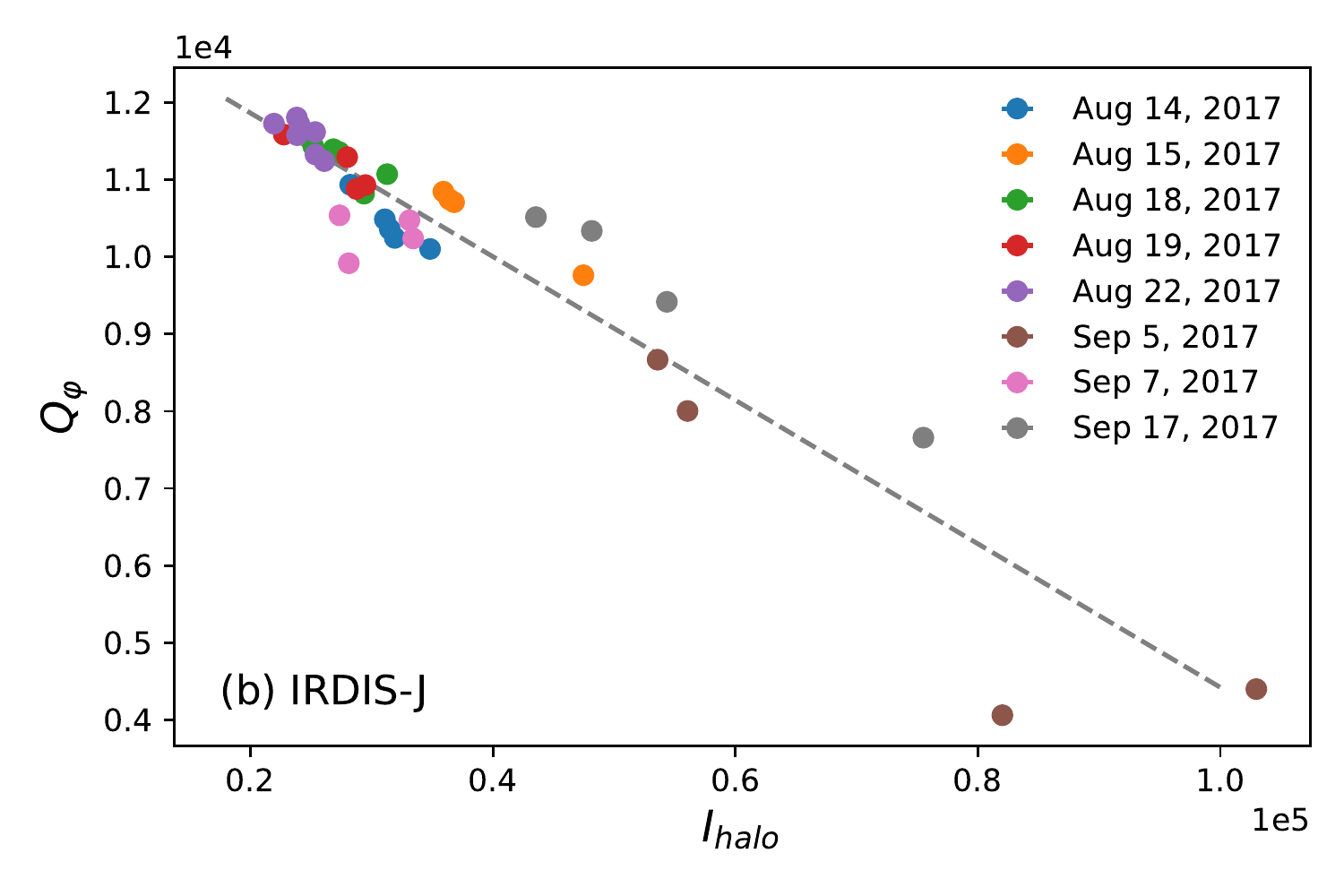}
        \label{fig:corr-qphi-halo-irdis-j}
    \end{subfigure}
    \begin{subfigure}[]{0.45\textwidth}
        \includegraphics[width=\textwidth]{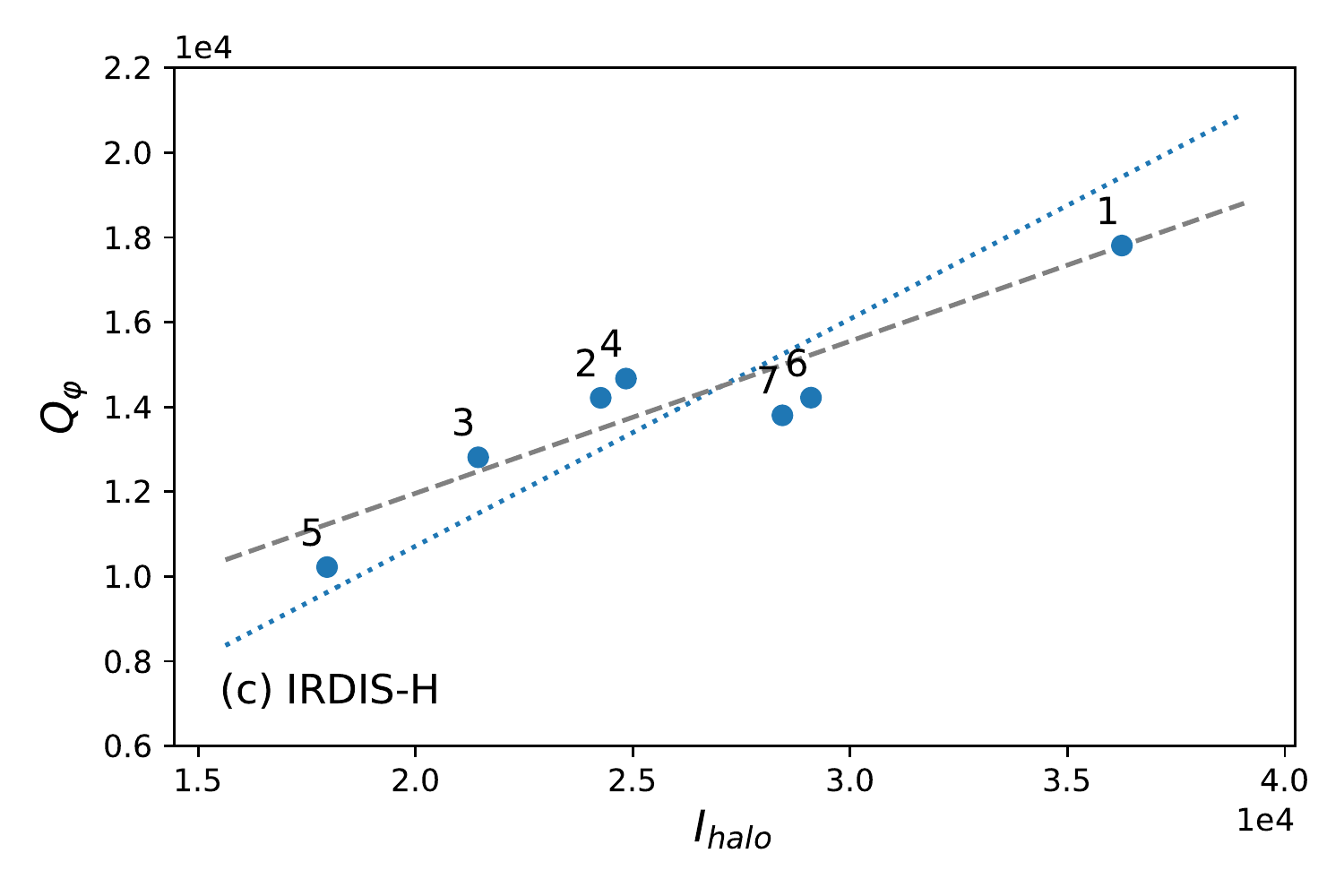}
        \label{fig:corr-qphi-halo-irdis-h}
    \end{subfigure}
    \vspace{-0.5cm}
    \caption{Dependences (gray lines) between the measured count rates for the polarized flux of the disk, $Q_{\varphi}$, of RX~J1604 and the stellar intensity in the halo, $I_{halo}$, for 
    (a) the six R band cycles from June 11, 2015, 
    (b) the 37 J band cycles from the eight nights in August and September 2017, and
    (c) the seven H band cycles from June 22, 2018. The dotted blue line in (c) represents a direct proportionality $Q_{\varphi} = 0.54 I_{\rm halo}$ as expected for transmission variations caused by clouds.}
    \label{fig:corr-qphi-halo}
\end{figure}

The variability of the PSF complicates the determination of the intrinsic polarization profile $\hat{Q}_\varphi(r)$ from the observations. Our archival data of RX~J1604 were taken as part of different observing programs and the measuring procedure was different for the R, J, and H bands. Therefore, we discuss the three data sets individually.

\paragraph{R band data.} For ZIMPOL, we have for each $Q_\varphi(x,y)$ map a simultaneous PSF. The PSF quality is characterized by peak flux ${\rm SB}_{\rm peak}$, which is defined as the average counts within the central 4 pixels and normalized to the solid angle of 
one milliarcsecond squared (mas$^{-2}$). We find for the R band data of RX~J1604 a strong correlation between PSF peak flux ${\rm SB}_{\rm peak}/I_{\star}$ and $Q_{\varphi}/I_{\star}$ as shown in Fig.~\ref{fig:corr-qphi-peak}{a} because polarimetric cancellation is stronger for a bad PSF or a low PSF peak flux as described in the analysis of \cite{Tschudi21}. 
This correlation explains a significant part of the dispersion obtained for the measured azimuthal polarization $\sigma(Q_{\varphi})$ in Table~\ref{tab:integrated-intensity-summary}. 

For each cycle one can therefore derive, according to the procedure described above, the best fitting, intrinsic disk profile, $\hat{Q}_{\varphi}(r)$, which takes the PSF effects properly into account \citep{Tschudi21}. 
It is therefore not surprising that the derived intrinsic azimuthal polarization $\hat{Q}_{\varphi}/I_\star$ for the R band given in Table~\ref{tab:fitting-parameters} has a much smaller relative dispersion or uncertainty $\sigma(\hat{Q}_{\varphi})/\langle \hat{Q}_{\varphi}\rangle=4~\%$ than for the observed value ${Q}_{\varphi}$ given in Table~\ref{tab:integrated-intensity-summary} because the model convolution and fitting procedure correct for the seeing-induced $Q_{\varphi}$ variability. This is a particularly important aspect of the simultaneous ZIMPOL measurements because an accurate R band polarization flux measurement can be achieved for RX~J1604 despite the rather bad and strongly variable AO performance. 

\paragraph{J band data.} For the J band we used the five high-quality runs for the determination of $Q_{\varphi}$ for the disk around RX~J1604. We see for these data sets a weak correlation between $Q_{\varphi}$ and the normalized peak flux (Fig. \ref{fig:corr-qphi-peak}b). However, this correlation is not well defined, because $Q_{\varphi}$ and ${\rm SB}_{\rm peak}/I_{\star}$ are not measured simultaneously for these coronagraphic observations. Because of short-term seeing variations, the values for the $Q_{\varphi}$ show a spread for an individual night that is about half the dispersion measured for the mean values from all five nights $\sigma(\langle Q_{\varphi}\rangle)$.
The lack of strictly simultaneous $Q_{\varphi}$ and PSF measurements introduces an enhanced uncertainty for the J band in the determination of the intrinsic disk polarization $\hat{Q}_\varphi/I_\star$ given in Table~\ref{tab:fitting-parameters} when compared to the R band. 

\paragraph{Correlations between halo intensity and polarization.}
\label{section:correlation}
The PSF variability is a problem for the quantitative polarimetry in the coronagraphic J band imaging because the observations of the PSF and the disk signal $Q_\varphi$ are not simultaneous.
Therefore, we investigate whether one can use the intensity halo of the central star instead of the central intensity peak for simultaneous monitoring of the PSF quality. It can be expected for a source with constant total brightness that the halo flux of the star in the coronagraphic image is low if the PSF peak behind the coronagraphic mask is strong and vice-versa. For the RX~J1604 data, we determine the halo flux $I_{\rm halo}$ by integrating $I_{\rm obs}(x,y)$ in the $1.0\arcsec<r<1.5\arcsec$ annulus aperture.

The non-coronagraphic ZIMPOL data clearly show the expected anticorrelation between $Q_{\varphi}$ and $I_{\rm halo}$ in Fig.~\ref{fig:corr-qphi-halo}{a,} which is almost exactly opposite to the correlation between $Q_{\varphi}/I_{\star}$ and ${\rm SB}_{\rm peak}/I_{\star}$ shown in Fig.~\ref{fig:corr-qphi-peak}{a}. A well-defined anticorrelation between $Q_{\varphi}$ and $I_{\rm halo}$ is also obtained for the coronagraphic observations in the J band as shown in Fig.~\ref{fig:corr-qphi-peak}{b}, which includes not only the 25 cycles taken during the five selected "good" nights but also the 12 cycles from the "bad" nights that were discarded because of bad seeing (September~5 and 17) or cloudy weather (September~7). 
The anticorrelation between $Q_{\varphi}$ and $I_{\rm halo}$ is better defined than the correlation with the $I_\star$ measurement shown in Fig.~\ref{fig:corr-qphi-peak}{b}, because of the simultaneous measurement of $Q_{\varphi}$ and $I_{\rm halo}$. Thus, the variation of the flux of the stellar halo in coronagraphic observations seems to be a good indicator for PSF variations of the data. Perhaps, this effect can be used to improve the correction of PSF smearing effects for coronagraphic data. 
However, this requires a more detailed investigation and understanding of the halo intensity in coronagraphic data as a function of the PSF quality. This is beyond the scope of this paper because we could obtain a quite accurate $Q_{\varphi}/I_\star$ value for RX~J1604 from the average of five "good" J band runs. 

Interestingly, the measured dependence for the H band run shows a correlation, not an anticorrelation, between $Q_{\varphi}$ and $I_{\rm halo}$, as plotted in Fig.~\ref{fig:corr-qphi-halo}{c}.
This completely different behavior is expected if cloudy weather leads to atmospheric transmission variations that equally affect the polarized flux of the disk and the flux of the star. 
The best fitting linear relation yields a flatter slope than the best fit $Q_{\varphi} = 0.54\, I_{\rm halo}$ 
assuming pure transmission variations,
most likely because of the contribution from the anticorrelation effect between $Q_{\varphi}$ and $I_{\rm halo}$ introduced by the fast PSF variations as observed for the R band and J band data. This example shows that the dependence between $Q_{\varphi}$ and $I_{\rm halo}$ is also useful to recognize and disentangle atmospheric PSF variations and atmospheric flux transmission variations.

\subsection{Wavelength dependence of the disk radius}
\label{sect:rlambda}
\begin{figure}
    \centering
    \includegraphics[width=0.48\textwidth]{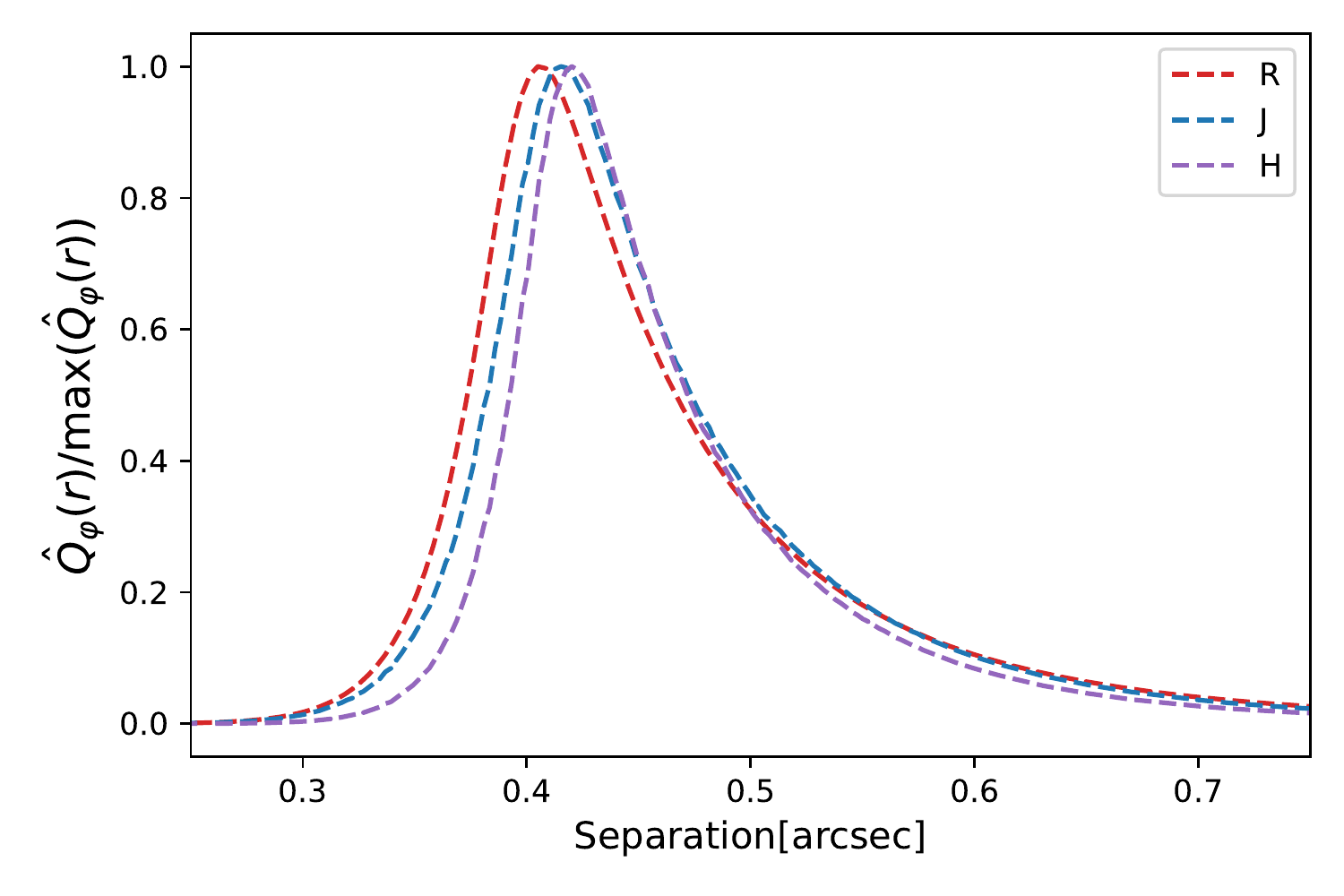}
    \caption{Intrinsic disk profiles, $\hat{Q}_\varphi(r),$ for RX~J1604 derived
    for the R band (red), the J band (blue), and the H band (purple).
    The profiles are normalized to their peak values.}
    \label{fig:rad-wavelength}
\end{figure}

We see in our multiwavelength analysis of RX~J1604 a small but systematic increase in the radius of the disk ring with wavelengths. This dependence is already described briefly in \citet{Pinilla2015} based on the comparison of the ZIMPOL data used also in this work and H band HiCIAO data from \citet{Mayama2012}. 
We can determine accurately the locations of the dust scattering surface of the illuminated inner disk wall for different wavelengths. It is found that the PSF convolution leads to an increase in the ring radius at the level of about 1~\% depending on the PSF structure. Therefore, we determined the ring radii for the peak flux from the best fitting intrinsic disk profiles, $\hat{Q}_\varphi(r),$ plotted in Fig.~\ref{fig:rad-wavelength} for the three wavelengths bands
and obtain the angular radii for the intrinsic peak flux of $r_{\rm peak}= 0.405\arcsec$, $0.416\arcsec,$ and $0.420\arcsec$ or physical radii of $60.9$~AU, $62.5$~AU, and $63.1$~AU for the R, J, and H bands, respectively. It should be noted that $r_{\rm peak}$ is different from the $r_0$ parameter used in the disk fits given in Table~\ref{tab:fitting-parameters}.

The difference between R band and H band disk radii is about 2.2~AU. This is an interesting value for a more detailed (future) investigation of the radial and vertical density structure and the distribution of dust particles in the inner wall of the disk and for the wavelength dependence of the dust absorption $\kappa(\lambda)$ and scattering $\sigma(\lambda)$ at this location.

\subsection{Disk intensity and fractional polarization}
\label{section:disk-intensity}
\begin{figure}[h]
    \centering
    \includegraphics[width=0.46\textwidth]{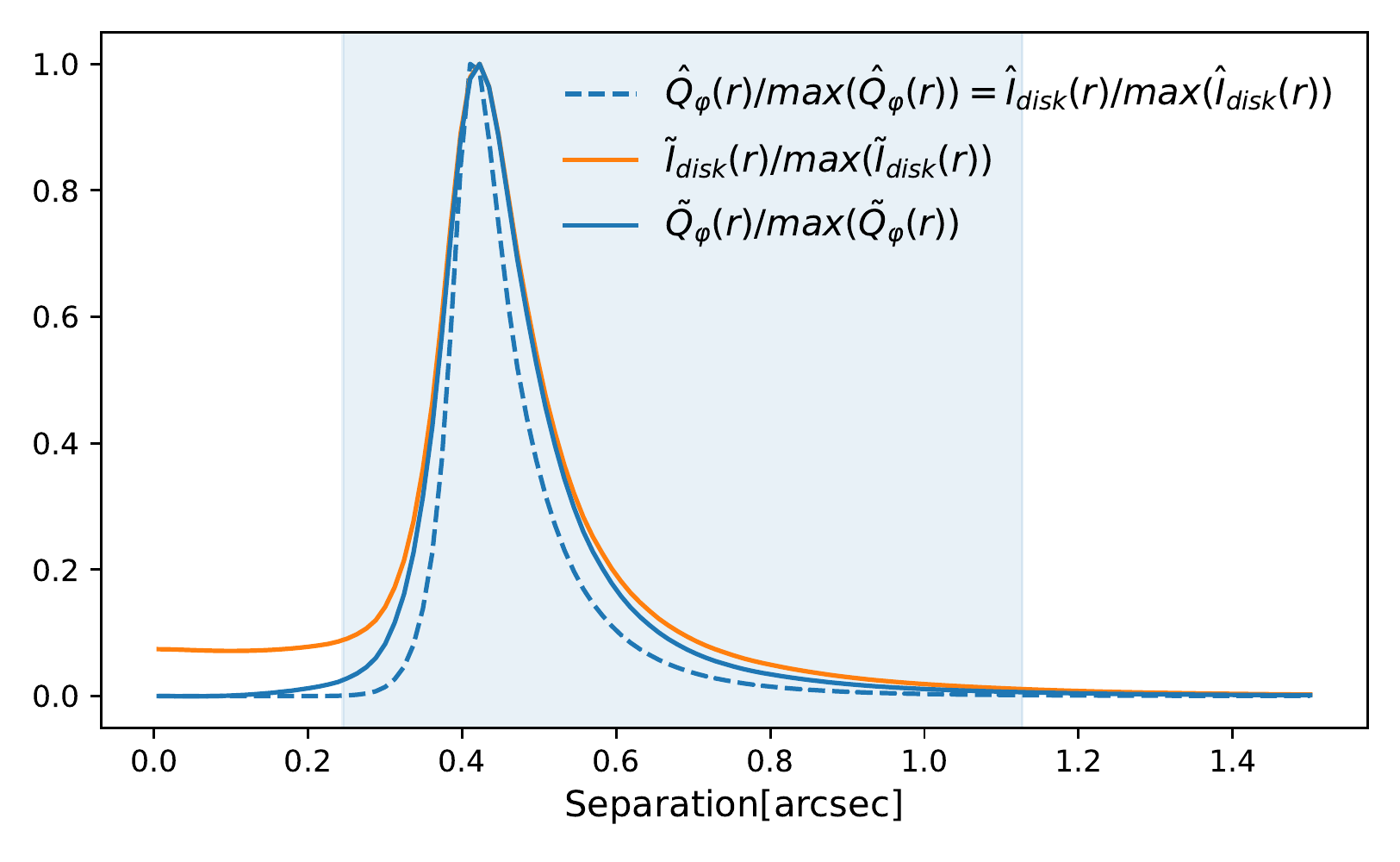}
    \caption{
    Radial profiles of RX~J1604 in the J band for the intrinsic disk intensity, 
    $\hat{I}_{disk}(r)$, for the convolved profiles  $\tilde{I}_{disk}(r)$ and
    $\tilde{Q}_\varphi(r)$. All profiles are normalized with respect to
    their peak value.}
    \label{fig:idisk_conv}
\end{figure}

\begin{figure*}
    \centering
    \begin{subfigure}[!ht]{0.47\textwidth}
        \includegraphics[width=\textwidth]{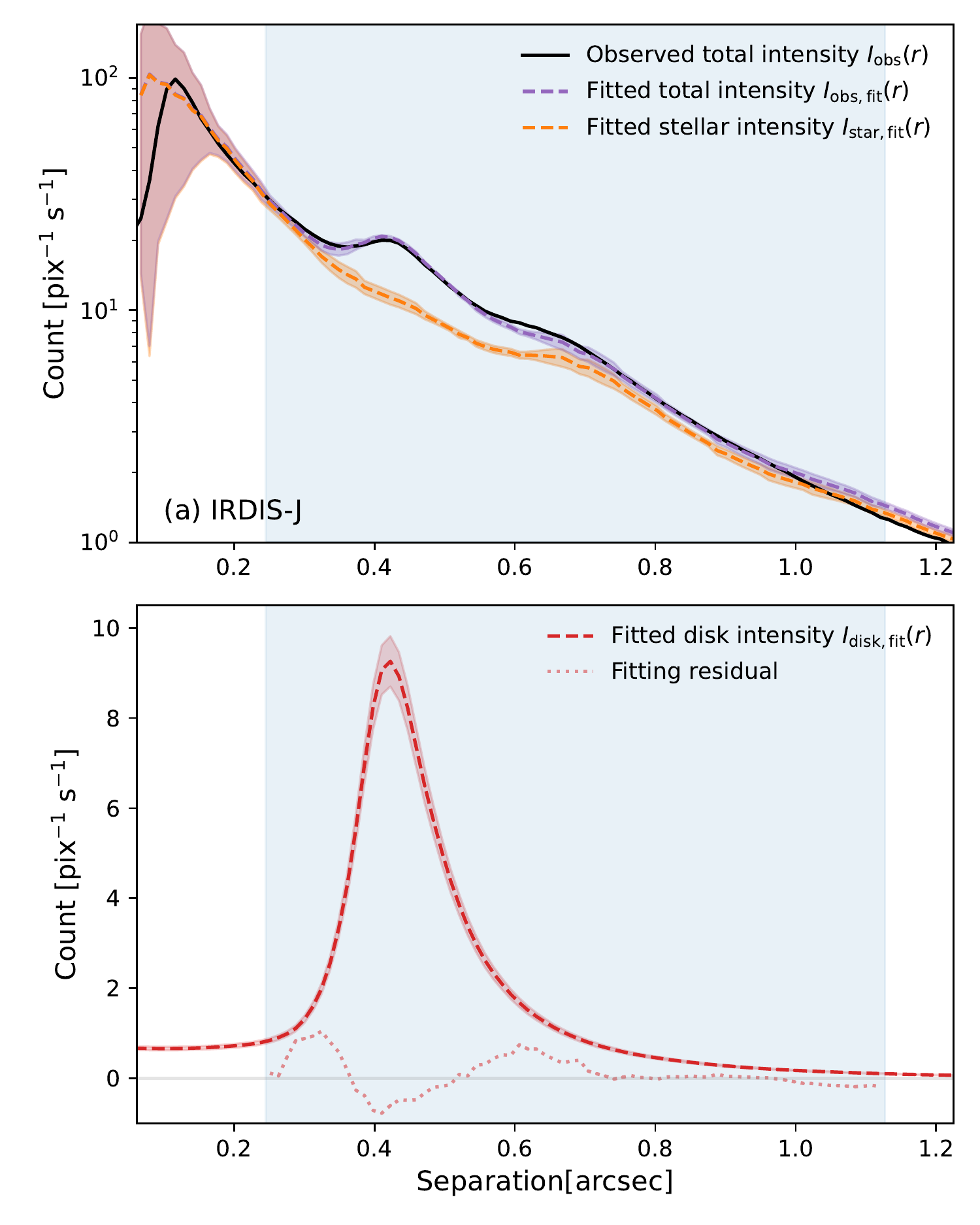}
        \vspace{-0.4cm}
        \label{fig:idisk_fit_j}
    \end{subfigure}
    \begin{subfigure}[!ht]{0.47\textwidth}
        \includegraphics[width=\textwidth]{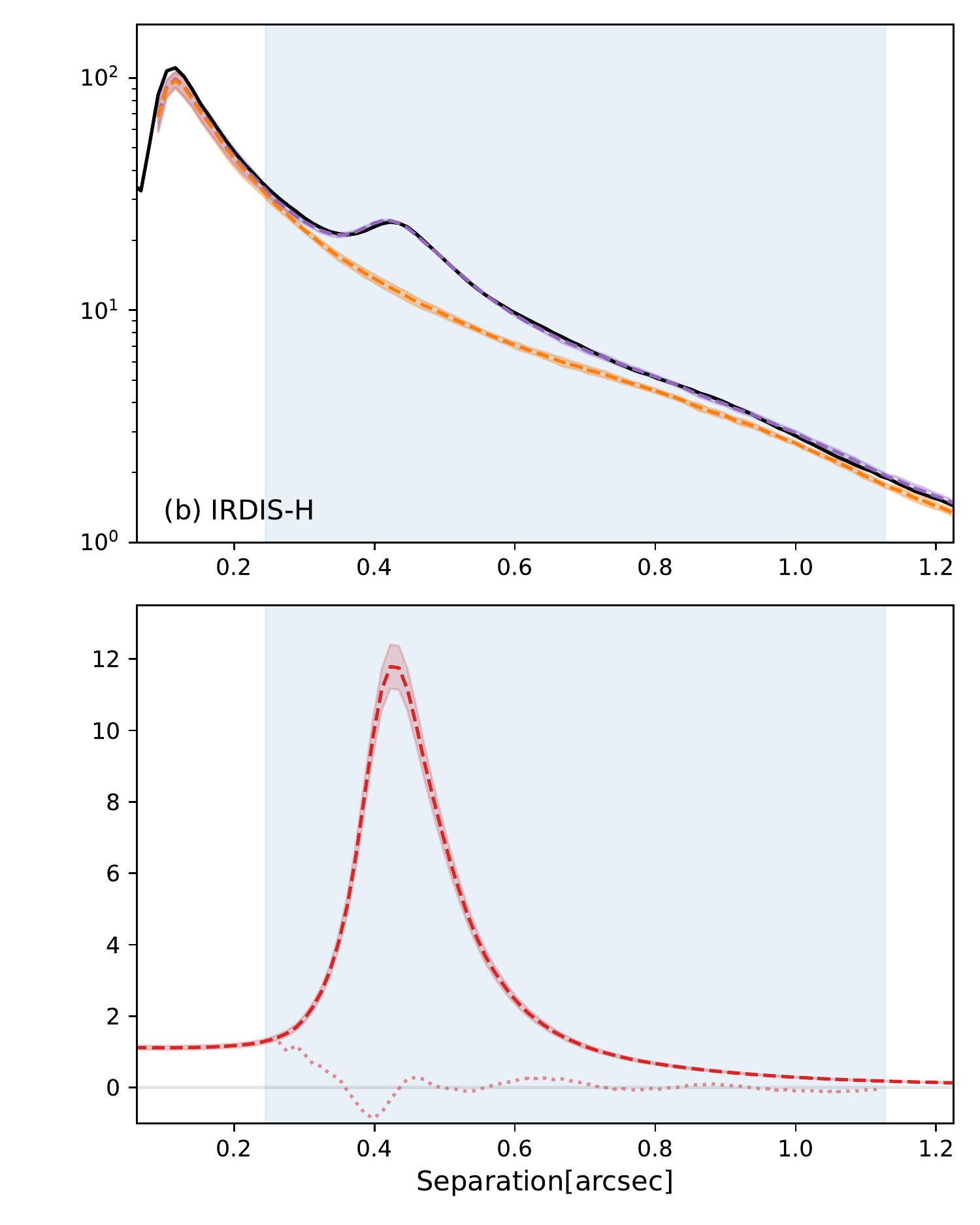}
        \vspace{-0.4cm}
        \label{fig:idisk_fit_h}
    \end{subfigure}
    \caption{Fitting result for the azimuthally averaged radial profile of disk intensity. Columns a and b are for the IRDIS-J and IRDIS-H band, respectively. Top panel: Radial profiles in logarithmic scale of the observed total intensity, $I_{\rm obs}(r),$ in black, the fitted total intensity, $I_{\rm obs, fit}(r),$ in purple, and the fitted stellar intensity, $I_{{\rm star, fit}}(r),$ in orange. Lower panel: Radial profiles of fitted disk intensity, $I_{{\rm disk, fit}}(r),$ in red and the fitting residual in brown.} 
    \label{fig:idisk_fit}
\end{figure*}

The intensity frames $I_{\rm obs}$ of RX~J1604 presented in Fig.\ref{fig:qphi-itot-image} include the stellar intensity and a weak contribution from the disk. In the R band, the location of the disk coincides with the ring of strong speckles defined by the AO control radius. Therefore, we were not able, and it seems impossible, to disentangle the disk intensity from the stellar signal in the R band data. 
Contrary to this, the peak brightness of the disk is located inside the speckle ring for the J band and the H band and can be recognized in the intensity data as clear bumps near $0.4\arcsec$ in the azimuthally averaged radial profiles presented in Fig.~\ref{fig:idisk_fit}. We extracted the disk intensity profile from the total intensity profile by fitting the stellar and disk intensity at the same time:
\begin{equation}
\label{eq:idisk-fit}
    I_{\rm obs}(r) = I_{\rm star}(r) + I_{\rm disk}(r)\,.
\end{equation}

\paragraph{Stellar intensity:} To simulate $I_{\rm star}$, we generated a set of $n$ radial profiles $I^n_{\rm ref}(r)$ containing $n$ coronagraphic profiles from deep polarimetric IRDIS observations of HD~150193 for the J band ($n=297$) and Hen3-1258 $(n=288)$ for the H band, taken with the same coronagraph and under similar atmospheric conditions as our RX~J1604 data. Only a few such data sets without strong polarimetric signals from a disk are available. Therefore, we extracted for each frame many profiles by averaging azimuthally over $\Delta \varphi = 10^\circ$ sections.

For the J band reference HD~150193A, there is a close companion B located at a distance $1.12\arcsec$ at position angle PA$=223^\circ$ \citep{Garufi14, Monnier17}, inside the field of view. To avoid the intensity contamination of the companion in the radial profile, we discard the southwest quadrant PA=[$180^\circ$, $270^\circ$]. For the remaining part of each image, we extract $\Delta \varphi = 10^\circ$ reference profiles $I_{\rm ref}$ at position angle $\varphi=5^\circ, 15^\circ, ..., 175^\circ, 275^\circ, 285^\circ, ..., 355^\circ$. 
In total, we obtain $n=27\times 11=297$ reference profiles for the J band. 

For the H band reference star Hen3-1258, we have four images and we extract reference profiles $I_{\rm ref}$ at position angle $\varphi=5^\circ, 10^\circ, ..., 350^\circ, 355^\circ$. We deliberately chose denser sections to boost the number of reference radial profiles, given that only four coronagraphic images are available for the H band. This results in $n = 72\times 4 =288$ reference radial profiles for the H band.

Our derivation of reference profiles is useful, but their quality could be improved with dedicated reference star calibrations taken during the same night or taken from a library. Unfortunately, no suitable data of this kind are available for the IRDIS polarimetry of RX~J1604. 

\paragraph{Disk intensity:} We adopt for the intrinsic intensity profile $\hat{I}_{\rm disk}(r)$ the same radial shape as for the polarization profile derived in the previous section according to 
\begin{equation}
\hat{I}_{\rm disk}(r) = \frac{\hat{Q}_\varphi(r)}{\hat{p}_\varphi}\,.
\label{eq:idisk}
\end{equation}
This implies that we assume $\hat{p}_\varphi(r)={\rm const}$ or no radial dependence of the fractional polarization for the scattered light from the disk. 
This simplification seems to be acceptable because the models of \citet{Ma2022} indicate rather small changes for $\hat{p}_\varphi(r)$ in the slope of the inner wall with $r$ when assuming everywhere the same type of dust. 
For example, the "0.5-reference model" for $i=7.5^\circ$ produces a polarization of about $\hat{p}_\varphi\approx 37~\%$ for a wall slope of $\chi \approx 30^\circ$ or steeper with respect to the mid-plane. A disk with the same dust can only produce a significantly different polarization of $\hat{p}_\varphi\approx 37~\%$ for a flat disk section with a slope of $\chi = 12.5^\circ$, as expected for the fainter disk surface farther out that is illuminated under grazing incidence.  
It seems unlikely that such a change from $30~\%$ at the bright inner disk edge to $37~\%$ for a low surface brightness region further out could be measured in our data because of the limited quality of our reference star calibrations.

Convolution of intensity $\hat{I}_{\rm disk}(x,y)\rightarrow \tilde{I}_{\rm disk}(x,y)$ also smears the radial profile, $\tilde{I}_{\rm disk}(r)$, but no cancellation happens, unlike for the $Q_\varphi$ signal as described in \citet{Tschudi21}. Figure~\ref{fig:idisk_conv} shows the differences between the convolved profiles $\tilde{I}_{\rm disk}(r)$ and $\tilde{Q}_\varphi(r)$, which are most obvious inside the ring $r<0.3\arcsec$, where $\tilde{I}_{\rm disk}(r)$ is clearly larger than zero. 

\paragraph{Best fits for $I_{\rm disk}$.} 
According to $I_{\rm obs}(r)=I_{\rm star}(r)+I_{\rm disk}(r)$ 
(Eq.\ref{eq:idisk-fit}), the fitting uses for the search of the best stellar profile,
\begin{equation}
I_{\rm star}(r) = k^i\, I_{\rm ref}^i(r)
,\end{equation}
where $I_{\rm ref}^i(r)$ are reference profiles and $k^i$ corresponding scaling factors. 
For the best disk profile $I_{\rm disk}(r)$ we adopt the shape of the intrinsic polarization profile $\hat{P}(r)=\hat{Q}_\varphi(r)=\hat{p}_\varphi \times \hat{I}_{\rm disk}(r)$ according to Eq.~\ref{eq:idisk}. 
$\hat{P}(r)$ is then convolved with the corresponding PSF (like an intensity signal, not like a polarization signal) to obtain $\Tilde{P}(r)$. We searched for a scaling parameter, $c_p$, in the disk profile
\begin{equation}
I_{\rm disk}(r) =c_p \tilde{P}_{\rm disk}(r) \,,
\end{equation}
where $c_p$ is is related to the intrinsic disk polarization $\hat{p}_{\varphi}= 1/c_p$. 

For each radial profile in the reference library $I_{\rm ref}^{i}(r)$, we fit the simulated total intensity profile to the observed profile with two constants $k^{i}$ and $c^i_p$ within the shaded fitting region $0.24\arcsec < r < 1.13\arcsec$ in Fig.~\ref{fig:idisk_fit} and assess the fitting result by the root-mean-square of the residuals. After fitting all reference profiles, we selected the best ten fittings, $i_{\rm best}$, and considered the average of the corresponding reference radial profiles as the best fitting stellar intensity profile,
\begin{equation}
    I_{{\rm star,fit}}(r) = \langle k^{i_{
    \rm best}}I_{\rm ref}^{i_{\rm best}}(r) \rangle\,,
\end{equation}
and derived the disk intensity profile:
\begin{equation}
    I_{\rm disk,fit}(r) = \langle c_p^{i_{\rm best}} \tilde{P}_{\rm disk}(r) \rangle\,.
\end{equation}

Best intensity fits are evaluated for most data. Not considered are the J band data from August 19 and the H band cycles 2 and 4, because the fitting solutions produce still large residuals. This is most likely caused by the not simultaneous PSF and ${I}_{\rm disk}(r)$ measurements. Figure~\ref{fig:idisk_fit} shows as examples for the selected fitting result the J band data from 2017-08-22 data and the first cycle of the H band data for the best fitting profile $I_{{\rm obs,fit}}(r) = \langle k^{i_{\rm best}}I_{\rm ref}^{i_{\rm best}}(r) + c^{i_{\rm best}}_p \tilde{P}_{\rm disk}(r) \rangle$ including 1-$\sigma$ standard deviation.

In Fig.~\ref{fig:idisk_fit}, $I_{\rm obs, fit}(r)$ coincides well with the $I_{\rm obs}(r)$. The disk intensity is well extracted within the fitting region. We integrated the disk intensity profile in the range $0<r<1.5\arcsec$ for all azimuthal angles, $\varphi$, to get the final $I_{\rm disk}$ for the J and the H band given in Table~\ref{tab:fitting-parameters}.

\paragraph{Fractional polarization and intrinsic disk intensity.} For the J band, we derive the disk-averaged intrinsic fractional polarization $\hat{p}_{\varphi}(J) = \hat{Q}_{\varphi}/I_{\rm disk} = 38 \pm 4 \%$ and disk intensity $I_{\rm disk}/I_{\star}(J) = 3.9\pm 0.5\%$ for the selected four nights, with uncertainty being the standard deviation of these four repeated measurements. It is not surprising that the J band measurement has small uncertainties because the observational condition was good for the selected measurement.

Although the observing conditions were highly variable for H band observations, we obtain stable $\hat{p}_\varphi(H) = 42 \pm 2\%$ based on the five selected cycles. The small uncertainty $\sigma(\hat{p}_\varphi)/\langle \hat{p}_\varphi \rangle = 5\%$ benefits from simultaneous disk intensity $I_{\rm disk}$ and polarized intensity $Q_\varphi$ measurements so that the fractional polarization $\hat{p}_\varphi=\hat{Q}_\varphi/I_{\rm disk}$ parameters are not strongly affected by the described atmospheric transmission variations.
Based on the fitting result, the disk intensity varies $\sigma(I_{\rm disk})/\langle I_{\rm disk} \rangle = 7\%$ between the selected 5 cycles. Adopting the $\sigma(I_{\star})/\langle I_{\star} \rangle = 15\%$ same as $Q_{\varphi}$, we estimate the reflected disk intensity $I_{\rm disk}/I_{\star} = 3.8 \pm 0.8 \%$. These measurements are summarized in Table~\ref{tab:fitting-parameters}.

\subsection{Scattered radiation and IR excess}
The dust in the transition disk of RX~J1604 not only scatters stellar radiation of the central source but also absorbs radiation and reemits this energy as thermal emission in the mid-IR and far-IR range. 
Comparing the scattered intensity $I_{\rm disk}/I_\star$ or polarized intensity $\hat{Q}_\varphi/I_\star$ with the thermal emission from the disk measured as far-IR excess $L_{\rm fIR}/L_{\rm star}$ can therefore be used for estimating at least roughly the apparent scattering albedo or the polarized scattering albedo of the disk \citep{Tschudi21}. 

The SED of RX~J1604 is modeled in \citet{Woitke19} and they obtain an IR-excess of $0.18 L_\odot$ for $\lambda > 6.72~\mu$m and a luminosity of the central star of $L_\star = 0.76 L_\odot$. 
The IR-excess due to the thermal emission of the disk resolved by the SPHERE observations is at a level of $L_{\rm fIR}/L_\star\approx 24~\%$. 

This can now be compared with the J band result for the reflected intensity from the disk, $I_{\rm disk}/I_\star$, derived in the previous section: 
\begin{equation}
    \Lambda_I = \frac{I_{\rm disk}/I_\star}{L_{\rm fIR}/{L_\star}}  \approx 0.16 \,.
    \label{Eq:appAlbedo}
\end{equation}
The double ratio for the polarized intensities given in Table~\ref{tab:fitting-parameters} and IR excess yield 
\begin{equation}
    \Lambda_\varphi = \frac{\hat{Q}_\varphi/I_\star}{L_{\rm fIR}/{L_\star}}  \approx 0.063 \,
\end{equation}
for the J band  and $\Lambda_\varphi=0.038$ for the R band. The interpretation of these double ratios in terms of apparent disk albedo should take the wavelength dependence of the scattered radiation into account so that one can compare the total, or wavelength-integrated, scattered radiation by the disk with the total absorbed energy for which we use the IR excess as a proxy. The J band value for $I_{\rm disk}/I_\star$ is a value near the peak of the irradiated flux, which should be at least roughly representative of the relative wavelength-integrated scattered flux.

Interestingly, the $\Lambda_I$ and $\Lambda_\varphi$ parameters for the apparent disk albedos of RX~J1604 are very similar to the values derived for HD~169142, another system with pole-on transition disk \citep{Tschudi21}.
Double ratios $\Lambda_\varphi$ for a larger sample of the circumstellar disk are compiled by \citet{Garufi2017} and they find for transition disks an average value of about $\Lambda_\varphi \approx 2.9~\%$. These should be considered as rough values because the data are heterogeneous, show a large scatter, and measuring uncertainties are quite unclear. Nonetheless, we can conclude that the $\Lambda_\varphi$ values derived by us for RX~J1604 are similar, but probably a bit higher when compared to the average of transition disks.

\subsection{Azimuthal dependence of the scattered disk light}
\label{section:azimuthal_variation}
\begin{table}[]
    \centering
    \caption{RX~J1604 model disk fitting parameters for the intrinsic azimuthal polarization dependence, $A(\varphi)$ (Eq.~\ref{eq:azimuthal}), derived for the epochs of the R band, J band, and H band observations.}
    \begin{tabular}{c c c c}
    \hline\hline
                  & R\tablefootmark{a} & J & H \\
    \hline
    $A_0$ $\times 10^{-7}$    & 5.25 $\pm$ 0.16   & 7.97 $\pm$ 0.05 & 9.77 $\pm$ 0.14\\
    $A_1$ $\times 10^{-7}$    & 2.69 $\pm$ 0.14   & 3.20 $\pm$ 0.21 & 3.50 $\pm$ 0.37\\
    $\varphi_1$ [deg]         & 35.6 $\pm$ 1.3    & 94.8 $\pm$ 0.9  & 82.1 $\pm$ 2.8\\
    $\sigma_1$ [deg]          & 29.9 $\pm$ 2.8    & 12.9 $\pm$ 1.1  & 24.1 $\pm$ 3.5\\
    $A_2$ $\times 10^{-7}$    & 1.28 $\pm$ 0.14   & 2.36 $\pm$ 0.14 & 5.21 $\pm$ 0.67\\
    $\varphi_2$ [deg]         & 207.0 $\pm$ 3.0   & 263.9 $\pm$ 1.3 & 249.4 $\pm$ 1.6\\ 
    $\sigma_2$ [deg]          & 44.2 $\pm$ 7.0    & 20.1 $\pm$ 1.6  & 12.1 $\pm$ 2.0\\
    \hline
    \end{tabular}
    \tablefoot{$A_0$, $A_1$, and $A_2$ are given as relative flux per pixel of 3.6~mas $\times$ 3.6~mas with respect to the total flux of the star. 
        \tablefoottext{a}{The R band values might be affected by a polarization calibration error (see Sect.~\ref{AppNormalization})}.} 
    \label{tab:shadow-fitting-parameters}
    
\end{table}
\begin{figure}[t]
    \centering
    \includegraphics[width=0.48\textwidth]{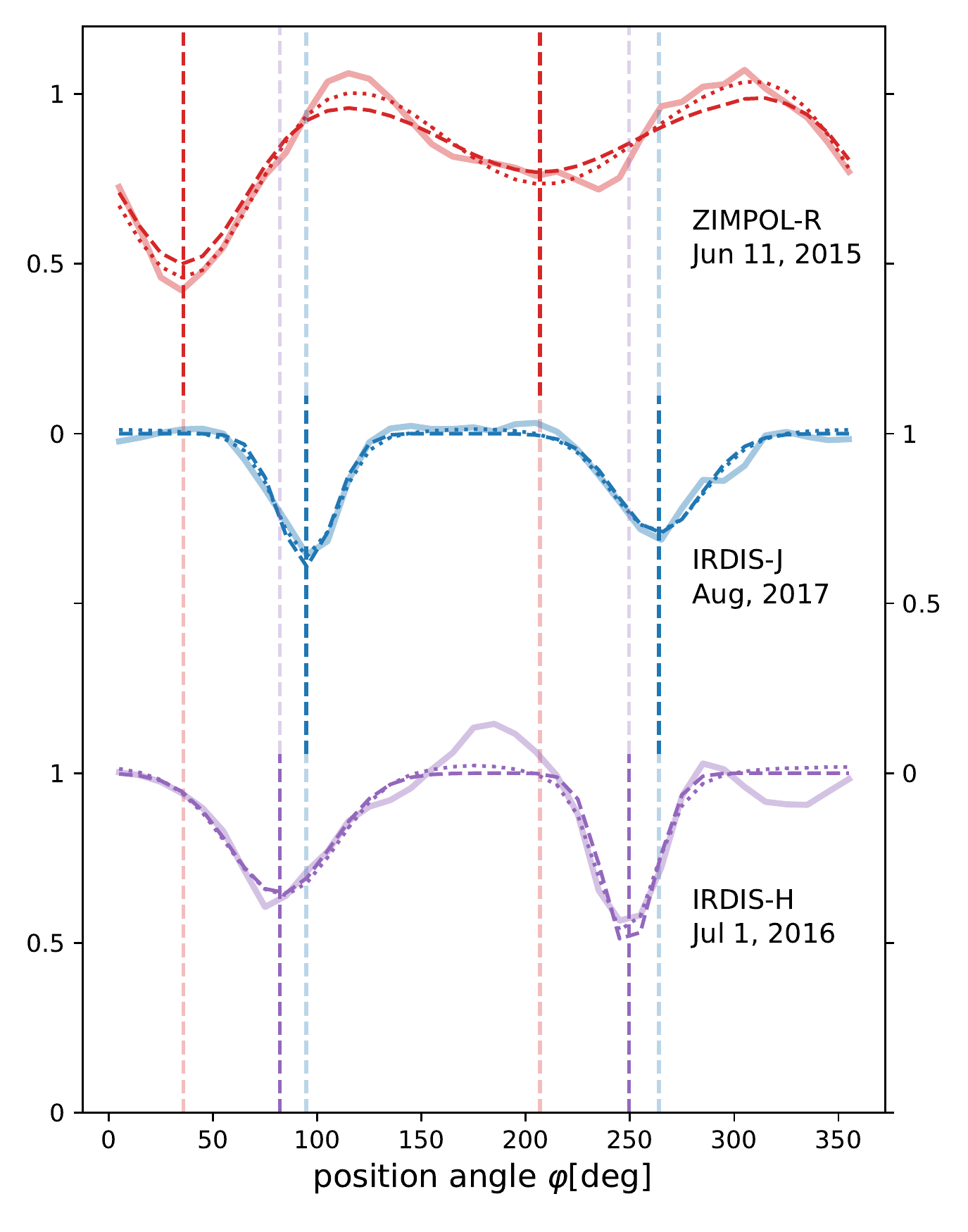}
    \caption{Observations (full line with light colors), fit (dotted lines), and convolved fit (dashed lines) of the azimuthal dependence of the polarized intensity, $A(\varphi)/A_0$, for RX~J1604. The curves give means of the R band epoch (top), of the five selected epochs for the J band (middle), and the epoch of the H band observations (bottom). The R band profiles might be affected by polarimetric calibration errors (see Sect.~\ref{AppNormalization}).}
    \label{fig:azimuthal-variance}
\end{figure}
The rotationally symmetric disk model adopted in the previous section for the determination of $\hat{Q}_\varphi$ and $I_{\rm disk}$ of RX~J1604 disregards the obvious shadows on the east and west side of the disk ring. The shadows in RX~J1604 are variable as described in \cite{Pinilla2018}, where they present the re-projected intensity for all epochs and determined the shadow locations and widths aiming to study the variance. In this section we characterize these shadows in terms of azimuthal brightness minima, with the goal to investigate the shadow effect on the disk intensity and polarization and determine the unattenuated disk intensity.

We introduce in the radial model of Eq.~\ref{eq: two-power} a position angle dependence $\hat{Q}_\varphi(r,\varphi)=(\hat{Q}_\varphi(r)/A_r)\cdot A(\varphi)$, where $A(\varphi)$ describes the shadows with Gaussians in the azimuthal brightness distribution: 
\begin{equation}
    A(\varphi) = A_0 - A_1\cdot \exp\left(-\frac{(\varphi - \varphi_1)^2}{2\sigma_1^2}\right)- A_2\cdot \exp\left(-\frac{(\varphi - \varphi_2)^2}{2\sigma_2^2}\right) \,.
    \label{eq:azimuthal}
\end{equation}
The amplitude $A_0$ describes the unattenuated disk polarization signal, while $A_1, A_2$, $\varphi_1, \varphi_2$, and $\sigma_1, \sigma_2$ give the depths, the azimuthal position angles, and the widths of the two shadows. We see the same radial brightness distribution $Q_\varphi(r)$ for bright and shadowed disk regions in our data. 

The best fitting solutions for the shadows are derived with a similar procedure as described in Sect. \ref{section:polarized-intensity-model}. We vary in the $\hat{Q}_\varphi(r,\varphi)$ model the parameters of the azimuthal dependence $A_\varphi(\varphi)$ in Eq.~\ref{eq:azimuthal}, convolve the models with the appropriate PSF and search for the best match between the convolved model profile $\Tilde{Q}_{\varphi}(\varphi)$ and the observed signal ${Q}_{\varphi}(\varphi)$. Both these azimuthal profiles are radially integrated between $0.18\arcsec<r<1.44\arcsec$. For this, the parameters $r_0, \alpha_{\rm in}$ and $\alpha_{\rm out}$ of the radial profiles were fixed as given in Table~\ref{tab:fitting-parameters}. The resulting parameters for the intrinsic profile $A(\varphi)$ are listed in Table~\ref{tab:shadow-fitting-parameters} for the averaged R, J, and H band observations. 
Figure~\ref{fig:azimuthal-variance} presents the observed azimuthal profiles $A(\varphi)$, the best fitting intrinsic model, and the convolved model radially integrated within $0.18\arcsec<r<1.44\arcsec$ for the different bands. The profiles are normalized to the unobscured sections of the disk and vertical dashed lines indicate the two shadow positions $\varphi_1$ and $\varphi_2$. 

We suspect that the polarimetric normalization of the data in the R band could introduce a significant error in the azimuthal profile $A(\varphi)$. Because of the low Strehl ratio of the R band data, the measured disk signal is weak with respect to the PSF of the central star, and therefore a small offset in the polarimetric calibration could introduce spurious structures as discussed in Sect.~\ref{AppNormalization}. In the J band and H band, the disk signal is much stronger than the stellar PSF, and therefore the derived azimuthal profile is much more robust. The following discussion will therefore concentrate mainly on the results of these two bands.

\paragraph{Interpretation of the shadows.}
The fit parameters $A_0$ for the unobscured disk ring sections (Table~\ref{tab:shadow-fitting-parameters}) are about 10~\% higher for the J band or 17~\% higher for the H band than the ring brightness $A_r$ obtained for the azimuthally averaged disk profile. Thus, the $A_0$ value should be used if one wants to deduce the reflectivity of disk sections not affected by shadows from the inner disk and this could also be the better choice for the derivation of dust scattering parameters. Assuming $A_r=A_0$ in Eq.~\ref{eq: two-power}, the polarized flux of the unattenuated disk would be $\hat{Q}_{\varphi}/I_{\star} = 1.63 \pm 0.08\%$ and $1.82\pm 0.53\%$ for the J and H band, respectively. This indicates a loss of about 7~\% and 16~\% for the illumination of the disk ring due to shadowing by the inner dust.

The azimuthal fits $A(\varphi)$ give different depths of the shadows for the R, J, and H bands. Unfortunately, the RX~J1604 data show quality differences and were taken at different epochs. It is not clear whether the strength of the shadows is a result of a wavelength-dependent absorption or of the temporal variability of the shadows as described in \citet{Pinilla2018}. 
For the temporal dust obscuration minima in the dipper light curves of RX~J1604, much stronger levels of dust absorption have been measured for shorter wavelengths, for example the V band when compared to the near-IR \citep{Sicilia2020}. This can be explained by optically thin absorption by small dust grains and it would be interesting to know, whether also the shadows on the transition disk of RX~J1604 show the same wavelength effect and could therefore be explained by the same type of optically thin dust. Multiwavelength shadow measurements taken during the same night would be useful to investigate this. 

Another interesting feature of the shadows on the RX~J1604 transition disk is the maximal depth of the azimuthal depression, which is less than 60~\% with respect to bright disk sections. The shadows from the inner disk reduce clearly the illumination, but even if the PSF convolution is considered, the shadow region is not completely dark and there remains at least 20~\% illumination compared to shadow-free disk regions.
Thus, the radiation of the central star is only partially reduced in the shadows, either because the absorbing dust is optically thin or because an optically inner dust disk only partly hides the central star. More detailed investigations of the shadows might provide interesting constraints about the geometry of the hot dust, but this is beyond the scope of this study. 

\paragraph{Fractional polarization.} We derived the azimuthal dependence of the polarized intensity $Q_\varphi(\varphi)$ and could also extract the scattered intensity of the disk $I_{\rm disk}(\varphi)$. Therefore, we investigated whether there exists an azimuthal dependence for the fractional polarization for the scattered light $p_\varphi(\varphi)$ and found for the J band no significant deviation from the mean value $\langle p_\varphi \rangle =38~\%$ exceeding the uncertainty range of $ \Delta p_{\varphi} = \pm 5~\% $. Therefore, we conclude that the fractional polarization $p_\varphi(\varphi)$ is constant within this range in agreement with model simulations of pole-on disks \citep{Ma2022}. 
We do not have enough signal in the shadows to identify a significant change in $p_\varphi(\varphi)$, which could be produced by the absorption and scattering of the hot dust near the star.

\section{Constraints on the dust scattering parameters}
\label{section:discussion}

An important goal of quantitative measurements of the scattered light from circumstellar disks is the derivation of dust scattering parameters, which can then be analyzed in terms of dust particle properties, like size distribution, structure, and composition. Therefore, we compare first the disk radiation parameters derived for RX~J1604 with simple fits to the model results from \citet{Ma2022} and try to understand the dependences between measured quantities and dust scattering parameters. 
The calculated grid of models is based on a simple, rotational symmetric transition disk geometry with an illuminated inner wall, described by three parameters for the inclination $i$ of the disk, the angular height $\alpha$, and the slope $\chi$ of the inner disk wall. The scattering by the dust is also described by only three parameters, the scattering albedo $\alpha$, the asymmetry parameter $g$ for the scattering phase function based on the Henyey-Greenstein function, and the scaled factor $p_{\rm max}$ for Rayleigh phase function defining the fractional polarization of the scattered radiation. The grid provides more than 50 combinations of model parameters the intensity and polarization images for the scattered radiation from which many observational parameters are derived, like the disk-integrated intensity $I_{\rm disk}/I_{\star}$, disk-integrated polarized intensity $\hat{Q}_{\varphi}/I_{\star}$, disk-averaged fractional polarization $\langle p_{\rm disk} \rangle$, maximum fractional polarization $\max(p_{\rm disk})$ and more. These model results are now compared with the measurements of RX J1604 presented in the previous section and used to constrain quantitatively the dust scattering parameters. The transition disk models of \citet{Ma2022} are particularly well suited for the disk around RX~J1604, because the model results for the low inclination case $i=7.5^\circ$ are very similar to the RX J1604 observations apart from the additional presence of the shadows on the east and west. However, for this analysis one needs to keep in mind  that the six parameter models of \citet{Ma2022} provide only a simple description of a real disk, in particular with respect to the description of dust scattering parameters.

In the second step, we used the model grid to search for the best fitting model for the derived radiation parameters of RX~J1604 for the J band. For this, we interpolated and extrapolated the calculated parameters of 11 models with $i=7.5^\circ$ and produced an extensive grid of model results, which we then used with an ensemble sampler based on the Markov chain Monte Carlo (MCMC) technique \citep{Foreman-Mackey2013}. 

\begin{figure}
    \centering
    \includegraphics[width=0.4\textwidth]{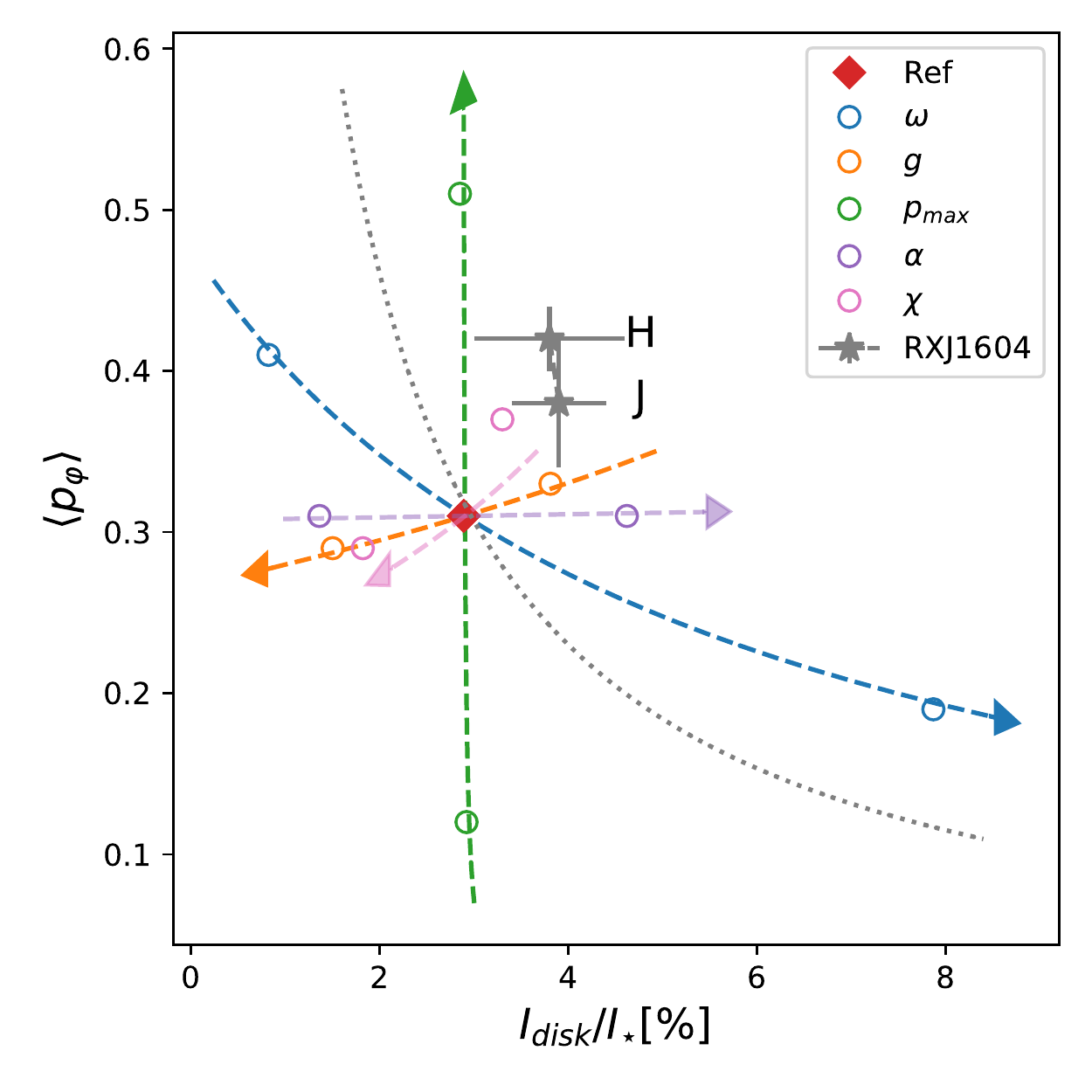}
    \caption{Radiation parameters for the disk in RX~J1604 in the fractional polarization $\langle p_\varphi \rangle$ vs. scattered intensity $I_{\rm disk}/I_\star$ plot. The gray data points show the J band and H band values and the dotted gray line the locations defined by $Q_\varphi/I_\star$ for the R band.
    Open points show results for transition disk models for $i=7.5^\circ$ from \citet{Ma2022}. The red point is the reference model with the parameters, $X_r$, given in Table~\ref{tab:diagnostic}, while the open points show models with one model parameter, $X_{r-1}$ or $X_{r+1}$, deviating from the reference case (values increase along the dashed arrows).\ The color coding is given in the inset.}
    \label{fig:param-grid}
\end{figure}

\subsection{Comparing RX~J1604 with selected model results}
The derived parameters for the reflected intensity $I_{\rm disk}/I_\star$, the polarized intensity $\hat{Q}_\varphi/I_\star$, and the averaged fractional polarization $\langle p_\varphi \rangle $ for the disk around RX~J1604 are summarized in Table~\ref{tab:fitting-parameters} for the three colors R, J, and H.
We compare these measurements in Fig.~\ref{fig:param-grid} with the results from 11 models of transition disks from Table~1 of \citet{Ma2022} for the inclination of $i=7.5^\circ$. The reference model is marked in red and the corresponding model parameters are scattering albedo $\omega=0.5$, scattering asymmetry $g=0.5$, maximum scattering polarization $p_{\rm max}=0.5$, disk wall height $\alpha=10^\circ$, and wall slope $\chi=32.5^\circ$ as listed in Table~\ref{tab:diagnostic} in column $X_r$. For the other models indicated in Fig.~\ref{fig:param-grid} by different colors one of the parameters differs from the reference case and has the value $X_{r-1}$ or $X_{r+1}$ given in Table~\ref{tab:diagnostic}. The arrows show how the disk radiation parameters change as the model parameters increases.

The derived parameters for the disk in RX~J1604 for the J band and H band lie on the high side for $I/I_\star$ and $\langle p_\varphi \rangle$ when compared to the selected model results. The J and H band values for $\langle p_\varphi \rangle$ are well defined, while the $I_{\rm disk}/I_\star$ values have larger uncertainties. 
For the R band we only have the $\hat{Q}_\varphi/I_\star=0.92~\%$ value. It is a coincidence that this value is almost exactly the same as the reference model value $0.90~\%,$ and therefore the dotted curve apparently passes through the red point. 

The J band reflectivity for RX~J1604, $I_{\rm disk}/I_\star=3.9~\%,$  is clearly higher than the reference model. The J band reflectivity would be even higher if one also considers the shadows of the disk. The unresolved, hot inner disk intercepts about 7\% of the stellar radiation, and without this shadowing the disk intensity would be about $I_{\rm disk}^{\rm corr}/I_\star=4.2\%$. 
Such a high disk scattering intensity could be explained by model parameters different from the reference case like a high scattering albedo $\omega>0.5$, a small forward scattering parameter $g<0.5$, a high disk wall height $\alpha>10^\circ$, or a rather flat wall slope $\chi<32^\circ$, as indicated by the colored arrows in Fig.~\ref{fig:param-grid}. Of course, one should consider combinations of these model parameters, and therefore the dependences of the disk radiation parameters on the model parameters are quite complex. However, it is possible to select certain observational parameters that better constrain dust scattering.

\begin{table*}[!ht]
    \centering
    \caption{Gradients, $A_{X_i}$, or functional dependence of the disk radiation parameters on disk model parameters.} 
    \begin{tabular}{c c c c c c c c c}
    \hline\hline
         & $X_r$ & $X_{r-1}$, $X_{r+1}$  
                & $I_{\rm disk}/I_\star$  
                    & $Q_\varphi/I_\star$ 
                        & $\langle p_{\varphi} \rangle$  
                            & $\Lambda_{I}$\\
     \hline
    $\omega$ & 0.5 & 0.2,0.8 
                &  $1.68\pm0.04$ \tablefootmark{a} 
                    & $1.10 \pm 0.04$
                        & $-0.59 \pm 0.03$
                            & $1.99 \pm 0.01$ \tablefootmark{a} \\
    $g$    & 0.5 & 0.25,0.75 
                & $-0.80 \pm 0.12$ 
                    & $-0.92 \pm 0.07$ 
                        & $-0.13 \pm 0.01$
                            & $-0.87 \pm 0.09$\\
                                    
    $p_{\rm max}$ & 0.5 & 0.2,0.8  
                & $-0.02 \pm 0.01$ 
                    & $1.01\pm 0.01$ 
                        & $1.03 \pm 0.02$
                            & $-0.02 \pm 0.01$\\
                                   
\noalign{\smallskip}    
    $\alpha$ & 10 & 5, 20
                & $1.13 \pm 0.05$ 
                    &  $1.13 \pm 0.04$
                        & $0.01 \pm 0.01$  
                            & $0.13 \pm 0.01$\\ 
   $\chi$   & 32.5 & 12.5, 57.5 
                & $-0.38 \pm 0.09$ 
                    & $-0.56 \pm 0.02$
                        & $-0.17 \pm 0.08$ 
                            & $-0.08 \pm 0.02$\\
    \hline
    \end{tabular}
    \tablefoot{
        The functional dependence of the integrated intensity $I_{\rm disk}/I_\star$, azimuthal polarization $Q_\varphi/I_\star$, disk-averaged fractional polarization $\langle p_\varphi \rangle$, and apparent albedo $\Lambda_I$ for the scattered and thermal light from the transition disk on the model parameters dust scattering albedo $\omega$, scattering asymmetry $g$, scattering polarization $p_{\rm max}$ and parameters for the disk geometry angular wall height $\alpha$ and wall slope $\chi$ (disk inclination is fixed to $7.5^\circ$).
        \tablefoottext{a}{The gradient is fitted using power law $Y/Y_r = \exp{[A'_X(X-X_r)/X_r]}$ because the linear relation does not fit well.} 
        }
    \label{tab:diagnostic}
\end{table*}

\paragraph{Fractional polarization.} It is clear from Fig.~\ref{fig:param-grid} that the disk-averaged polarization, $\langle p_\varphi \rangle$, of a disk depends mainly on the scattering polarization, $p_{\rm max}$, and the albedo, $\omega$, while the other model parameters have only a small impact. This is a specific property for pole-on disks, like RX~J1604, because the scattering angle $\theta_{s}$ is everywhere close to $\theta_s \approx 90^\circ$, where the produced scattering polarization is close to maximal for given dust scattering parameters. 
Therefore, $\langle p_\varphi \rangle$ is practically equal to the maximum fractional polarization max$(p_\varphi)$, and we can use the approximate relation between $p_{\rm max}$, $\omega$, and $\langle p_\varphi \rangle$ according to \citet[][Eq.~17]{Ma2022}:
\begin{equation}
\langle p_{\varphi} \rangle \approx p_{\rm max}\,(1.02 -0.76\cdot \omega)\,.
\label{Eq:pmax}
\end{equation}
This relation is largely independent of other model parameters. Smaller 
$\langle p_{\varphi} \rangle$ for larger $\omega$ is caused by the fact that high albedo dust produces a lot of multiple-scattered photons with randomized polarization, and thus reduces the strong polarization signal from the photons undergoing only one single scattering. 

Equation~\ref{Eq:pmax} defines for the derived J band value of $\langle p_{\varphi} \rangle=0.38$ a simple relation between $\omega$ and $p_{\rm max}$. This defines the lower limit $p_{\rm max}>0.37$ for $\omega\rightarrow 0$ when only single scatterings contribute to the scattered light. Another extreme case is by definition $p_{max}\rightarrow 1$ and this yields for the albedo the limit $\omega<0.84$. This relation also indicates that the $p_{\rm max}$ and $\omega$ can only be both low, intermediate, or high, for example, $p_{\rm max}=0.44$ for $\omega=0.2$, or $p_{\rm max}=0.92$ for $\omega=0.8$.

\begin{figure}
    \centering
    \includegraphics[width=0.45\textwidth]{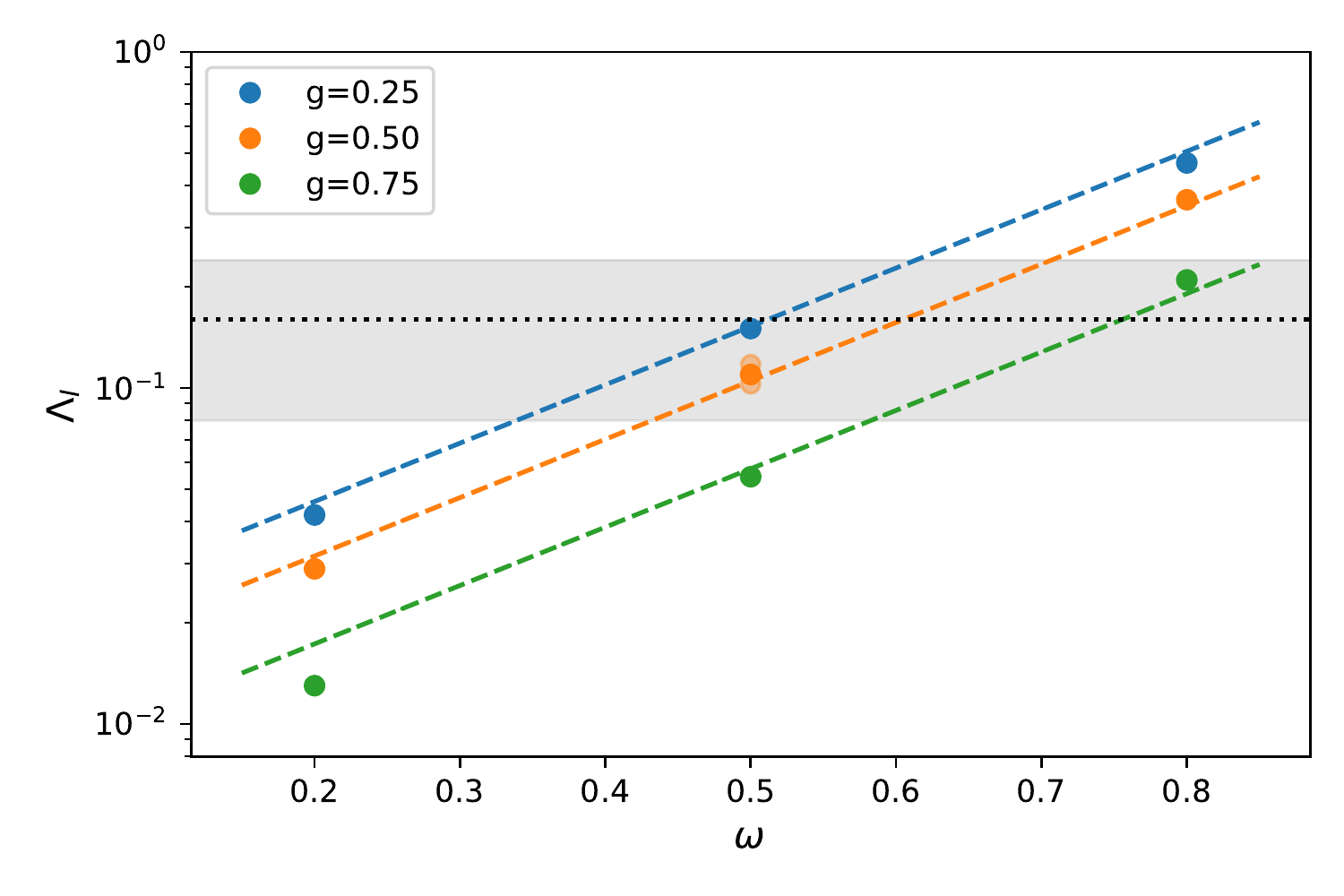}
    \caption{Apparent disk scattering albedo, $\Lambda_I$, as functions of the scattering albedo, $\omega$, and the asymmetry parameter, $g$ 
    (for $i=7.5^\circ$, $\alpha=10^\circ$, $\chi=32.5^\circ$, and $p_{\rm max}=0.5$). The dots are model results from \cite{Ma2022}, and the dashed lines represent the fit given in Eq.~\ref{Eq:LambdaI}. The horizontal line shows the derived value for the apparent albedo, $\Lambda_I=0.16\pm 0.08,$ for the disk in RX~J1604, and the shaded area indicates the uncertainty range.}
    \label{fig:lambdai_eq}
\end{figure}

\paragraph{Scattering and IR excess.} 
The apparent disk albedo $\Lambda_I$ (Eq.~\ref{Eq:appAlbedo}), defined as the double ratio between scattered intensity $I_{\rm disk}/I_\star$ and the far-IR thermal excess $F_{\rm fIR}/F_{\rm star}$, is another powerful disk radiation parameter for constraining the dust scattering. 
The ratio between scattering and absorption depends strongly on dust properties. In general, $\Lambda_I$ depends also on the disk geometry. But for a pole-on disk, these effects are small so that $\Lambda_I$ depends according to \citet{Ma2022} mainly on $\omega$ and $g$. as shown in Fig.~\ref{fig:lambdai_eq}. The points can be fitted for $i=7.5^\circ$, $\alpha=10^\circ$, $\chi=32.5^\circ,$ and $p_{\rm max}=0.5$ by a simple relation between $\Lambda_{I}$, $\omega,$ and $g$:  
\begin{equation}
    \Lambda_{I} \approx (0.20-0.19\,g)\cdot {\rm e}^{2(2\omega-1)} \,.
    \label{Eq:LambdaI}
\end{equation}
Of course, a high single scattering albedo $\omega$ for the dust also produces a high apparent disk albedo $\Lambda_I$, while a high $g$ parameter or strong forward scattering lowers the disk albedo because the photons are scattered predominantly deeper into the absorbing disk surface. 

For pole-on disks, $\Lambda_I$ is almost independent of the slope of the illuminated inner disk wall $\chi$ and does not depend significantly on the wall height $\alpha$ and on the scattering polarization $p_{\rm max}$. 
Therefore, the fit given in Eq.~\ref{Eq:LambdaI} or shown in Fig.~\ref{fig:lambdai_eq} is a good approximation for low inclination disks $i\leq 10^\circ$, with model parameters $\alpha\leq 20^\circ$ and $\chi \leq 57.5^\circ$ for the disk geometry, and $0.2\leq\omega\leq 0.8$, $0.2\leq p_{\rm max}\leq 0.8$, and $0.25\leq g \leq 0.75$ for the dust scattering.  

According to Fig.~\ref{fig:lambdai_eq}, the derived $\Lambda_I=0.16$ for RX~J1604 would be compatible with dust scattering parameter combinations $g\approx 0.21$ for $\omega=0.5$, or $g\approx 0.46$ for $\omega=0.6$, or $g\approx 0.80$ for $\omega=0.8$.  
This derivation assumes that the scattered intensity $I_{\rm disk}/I_\star$ or the apparent albedo $\Lambda_I$ for the J band is representative of the disk reflectivity $S$ and absorption $1-S$ for all wavelengths. 
The J band is at the peak of the stellar SED for RX~J1604 \citep{Woitke19}. We also measure a strong wavelength dependence for the polarized intensity, $\hat{Q}_{\varphi}/I_\star$, of the disk between the J band and R band of a factor of 1.64, and the $\Lambda_I$ parameter between the J band and the R band most likely behaves similarly.
Therefore, adopting J band as a representative is a rough assumption and we account for a relative uncertainty of $\pm 50~\%$ for $\Lambda_I$.
But even with these large uncertainties, $\Lambda_I$ constrains the $\omega$ and $g$ parameter rather well because of the exponential dependence of $\omega$.

\paragraph{Relation between scattering polarization and asymmetry.}
Equation Eq.~\ref{Eq:pmax} for $p_{\rm max}$ and $\omega$ and Eq.~\ref{Eq:LambdaI} for $g$ and $\omega$ can be combined into a relation between the scattering polarization, $p_{\rm max}$, and asymmetry, $g$, which takes the form
\begin{equation}
    g= 1.05 - 0.18\, \Lambda_I\cdot e^{5.26\langle p_\varphi \rangle/p_{\rm max}}\,.
\label{eq:gpmaxgeneral}   
\end{equation}
Inserting the measured values $\Lambda_I=0.16$ and 
$\langle p_\varphi\rangle = 0.38$ yields 
\begin{equation}
    g= 1.05 - 0.029 \cdot e^{2.00/p_{\rm max}}\,.
\label{eq:gpmax}
\end{equation}
According to Eq.~\ref{Eq:pmax}, the scattering polarization, $p_{\rm max}$, is restricted to $p_{\rm max}>\langle p_\varphi \rangle \approx 0.38$. Therefore, the combined equation for $p_{\rm max}$ and $g$ (Eq.~\ref{eq:gpmax}) yields only reasonable $g$ parameters for high $p_{\rm max}$ values; for example, $p_{\rm max}=0.6$, 0.7, or 0.8 yield $g=0.24$, 0.55, or 0.70, respectively. 
The range for $p_{\rm max}$ is less restricted if the error range for $\sigma(\Lambda_I)=\pm 0.08$ and $\sigma(\langle p_\varphi\rangle)= \pm 0.04$ are also taken into account.
The relation given in Eq.~\ref{eq:gpmaxgeneral} can also be applied for other pole-on disks with $\Lambda_I$ and $\langle p_\varphi\rangle$ measurements.

\begin{figure}
    \centering
    \includegraphics[width=0.49\textwidth]{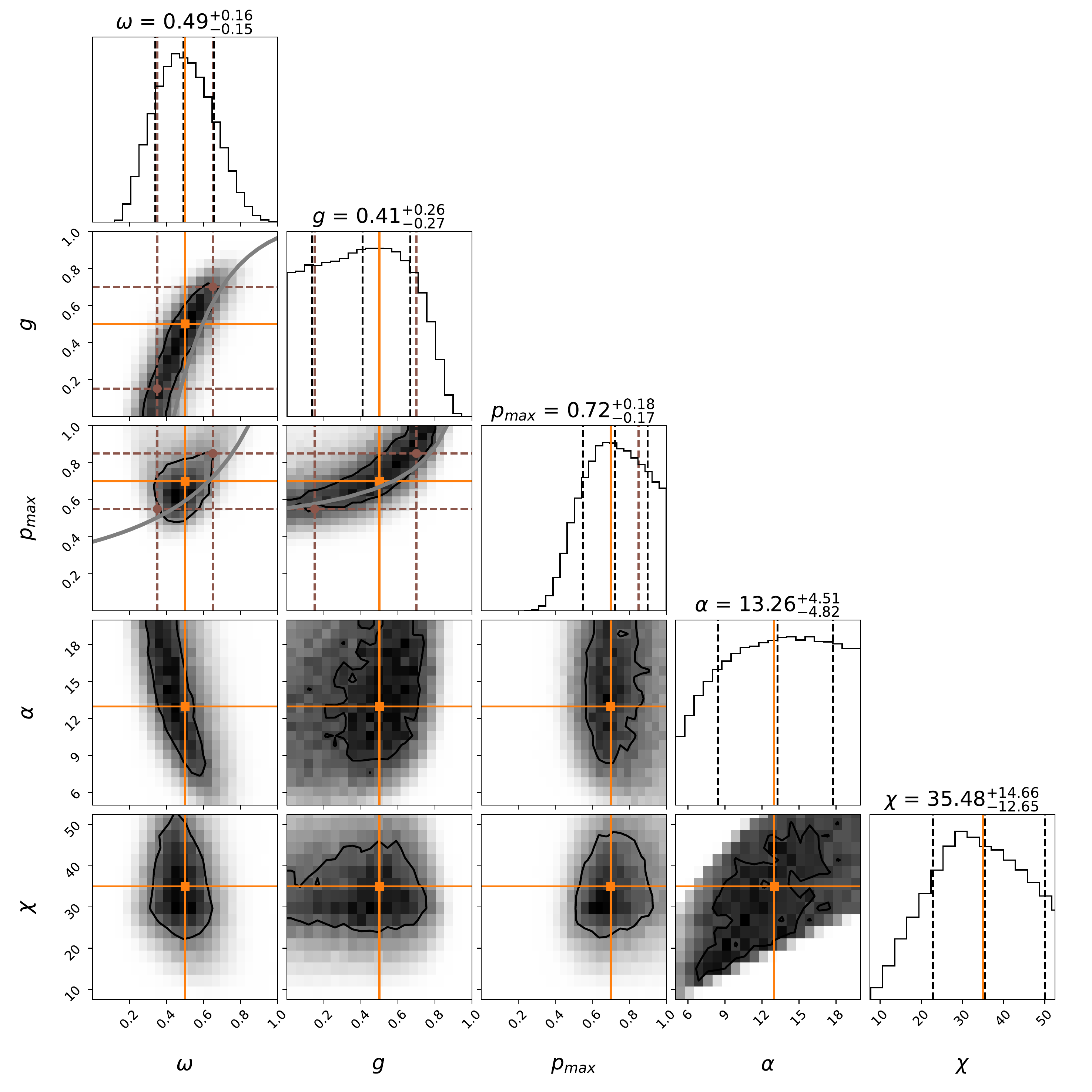}
    \caption{Posterior probability distribution for the scattering parameters $\omega, g$, and $p_{\rm max}$ and the geometric parameters $\alpha$ and $\chi$. The black contours show the $1\sigma$ level.}
    \label{fig:corner}
\end{figure}

\subsection{Model fitting with an MCMC method}
\label{section: mcmc}
In this section we compare the obtained model results for the intrinsic disk radiation parameters $I/I_\star$, $\hat{Q}_\varphi/I_\star$ and $\Lambda_I$ for the J band and derive the likelihood of the model parameters $\omega$, $g$, $p_{\rm max}$, $\alpha$ and $\chi$ (the disk inclination is fixed to $i=7.5^\circ$) using the standard MCMC ensemble sampler from \citet{Foreman-Mackey2013}.

For this, we needed model results for many model parameter combinations, $X=(X_1, X_2,\ldots),$ which can be obtained from the results of \citet[][Sect.~5.2]{Ma2022} with extrapolations of radiation parameters of the reference model $Y_r$ (e.g., the red point in Fig.~\ref{fig:param-grid}) from $X_{i,r}$ to $X_i$ via a product of linear relations:
\begin{equation}
    \frac{Y(X)}{Y_r} = \prod_{i}\,\, \Bigl(A_{X_i}\frac{X_i-X_{i,r}}{X_{i,r}}+1\Bigr)\,.
    \label{Eq:extrapolation}
\end{equation}
Here $X_{i,r}$ are the model parameters and $Y_r$ the disk radiation parameter for the reference case, and $A_{X,i}$ is the relative gradient of a linear relation $Y=aX_i +b$ at the reference point $(Y_r,X_r)$. The derived gradients $A_{X_i}$ are given in Table~\ref{tab:diagnostic} and they are determined from fits to the available result for the parameters $X_{r-1}, X_{r}$ and $X_{r+1}$. The dashed arrows in Fig.~\ref{fig:param-grid} are calculated from the extrapolations used in Eq.~\ref{Eq:extrapolation} and the agreement with model results illustrate the quality of this approach.
It should be noted that the dependence between $I/I_\star$ and $\omega$ cannot be well described by a linear relation and therefore an exponential function is used instead, as described in \citet{Ma2022}. The gradients $A_{X_i}$ given in Table~\ref{tab:diagnostic} are derived for the fixed disk inclination of $i=7.5^\circ$ and slightly differ from the results in Table~2 of \citet{Ma2022}.

We can determine with these procedures a dense grid of simulated radiation parameters $I_{\rm disk}/I_{\star}$, $Q_{\varphi}/I_{\star}$ and $F_{IR}/F_\star$ and carry out a maximum likelihood search with MCMC using uniform priors and the disk radiation parameters $I_{disk}/I_{\star} = 3.9\pm 0.5\%$, $Q_{\varphi}/I_{\star}=1.51\pm 0.11 \%$, and the apparent disk albedo $\Lambda_I = 0.16 \pm 0.08$ of RX~J1604 for the J band from Table~\ref{tab:fitting-parameters}. 
For the dust scattering, the selected parameter ranges are $\omega =(0,1)$, $g=(0,1)$, $p_{max}=(0,1)$ to cover all possible values. For the geometric parameters, the range of $\alpha=(5^\circ, 20^\circ)$ is chosen as the typical values from observations \citep[e.g.,][]{Rich2021,Avenhaus2018}.
The angles $\chi$ and $\alpha$ are related by $\tan\chi = \tan\alpha \cdot r_w/(r_w-r_i)$, where $r_i$ and $r_w$ are the inner radius and the rim radius of the wall, respectively. The slope $\chi$ is computed using the range $r_w/(r_w-r_i)=(1.5, 4)$ estimated from the observed disk ring width.

\paragraph{Best fitting model parameters.} The resulting 1D and 2D posterior probability distributions for the model parameters are shown in Fig.~\ref{fig:corner}. The 1D distributions define the best values and the $\pm 1 \sigma$ limits. 
They constrain the scattering albedo to $0.34<\omega<0.65$, the asymmetry to $0.14<g<0.67$, and the polarization to $0.55<p_{\rm max}<0.90$. 

The 2D distributions provide for a given scattering parameter two best values depending on the pairing of parameters, for example, $\omega-g$ and $\omega-p_{\rm max}$. Therefore, we selected a trio of best values for the dust scattering parameters $(\omega=0.50, g=0.50, p_{\rm max}=0.70)_{\rm best}$, which closely approximate all the peaks in the 2D distributions as shown by the orange points in the Fig.~\ref{fig:corner}.
There are strong correlations or narrow bands in the 2D probability distributions between $\omega$ and $g$ and $g$ and $p_{\rm max}$ and this defines in the 3D space $(\omega,g,p_{\rm max})$ a narrow band of specific combinations from low ($-1\sigma$) values $(0.35,0.15,0.55)$, to the peak value given above and to high ($+1\sigma$) value combinations $(0.65,0.70,0.85)$.

The best fitting model, $(0.50,0.50,0.70)_{\rm best}$, gives the radiation parameters $\hat{Q}_\varphi^{\rm fit}/I_\star = 1.62~\%$, $\langle \hat{p}_\varphi^{\rm fit}\rangle =0.43$, and $\Lambda_I^{\rm fit}=0.10$, which are values within $\pm 1\sigma$ of the measured values. The measured value for the apparent disk albedo with $\Lambda_I=0.16$  is  clearly higher, and therefore the fit from Eq.~\ref{Eq:LambdaI} using this $\Lambda_I$ value, shown in the $\omega-g$-panel in Fig.~\ref{fig:corner} as a gray line, is at about 1~$\sigma$ higher $\omega$ values when compared to the 
2D probability distribution. 
Also, the other fits described by Eqs.~\ref{Eq:pmax} for $\omega-p_{\rm max}$ and \ref{eq:gpmax} for $g-p_{\rm max}$ using the measured values (gray lines) are slightly offset because the Bayesian method combines the constraints from different measurements into one best solution compatible with all the measurements.
Nonetheless, this comparison shows that the simple fits provide already very reasonable approximations to the probability distributions based on a many-parameter MCMC fitting method.

For the disk geometry, the best fitting value for the angular wall height is $\alpha\approx 13^\circ$ and the wall slope $\chi\approx 35^\circ$. The peak of the 2D posterior distribution is well centered in the plotted $\alpha-\chi$ area but looks extended because of the restricted range for the priors. 
There is an anticorrelation between $\alpha$ and $\omega$ because the same intensity $I_{\rm disk}$ can be produced by a thin disk ($\alpha\approx 9^\circ$) with high albedo dust $\omega\approx 0.55$, or by a disk with low albedo dust, $\omega\approx 0.4,$ and a larger height, $\alpha\approx 18^\circ$, compensating for the lower reflectivity per surface element with a larger reflecting surface. This ambiguity is a fundamental problem for the interpretation of the intensity or polarized intensity, which is hard to resolve for pole-on disks.

\paragraph{Radial profile for the best fit.} The MCMC method was only applied to measurements of disk-integrated or disk-averaged radiation parameters without considering the measured radial profile $\hat{Q}_{\varphi}(r)$. Neglecting this important information is not optimal, but considering it would also require a grid of scattering models with a better description of the disk structure than just an inner wall with a constant slope and an angular height for the wall rim as adopted for the model grid used in this paper. Future model fitting should use more sophisticated disk geometries and include the radial profile in the analysis. 

However, we can still compare the best-fit solution with the observed profile $Q_\varphi(r)/I_\star$. The calculated disk-integrated radiation parameters do not depend on the radial extend, because the disk geometry in the scattering models of \cite{Ma2022} is characterized by the angular height $\alpha$ and wall slope $\chi$. 
Thus, the same amount of photons interact with the disk with given $\alpha$ and $\chi$ independent of the inner disk radius $r_i$. The two angles $\alpha$ and $\chi$ define a relation between the inner radius and rim radius of the wall $r_w$ according to $\tan\chi = r_w/(r_w-r_i)\cdot \tan\alpha$. 
With this relation, we can search for the inner radius, $r_i$, that yields for the best fitting model according to the MCMC method, a good match with the observed radial profile. Of course, the calculated intrinsic disk profiles must be convolved with the PSF of the observations $\hat{Q}_\varphi^{\rm fit}(r)/I_\star$ as described in Sect.~\ref{section:polarized-intensity-model} for the power law fitting of the radial profile.

\begin{figure}
    \centering
    \includegraphics[width=0.48\textwidth]{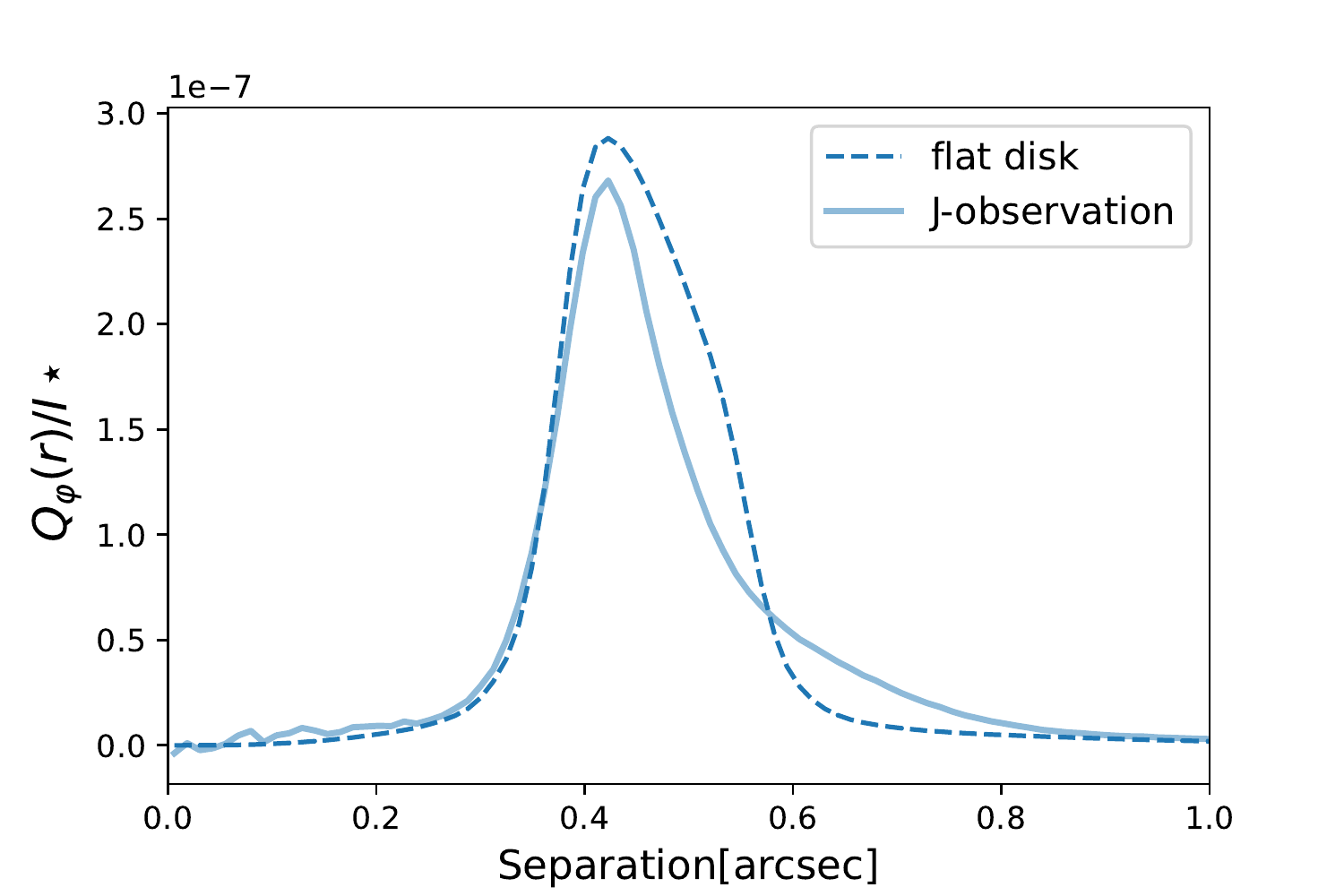}
    \caption{Convolved radial profile, $\tilde{Q}^{\rm fit}_{\varphi}(r)/I_{\star}$, based on the best fitting grid model according to the MCMC sampling (Fig.~\ref{fig:corner}). The disk parameters are $\omega=0.5$, $g=0.5$, $p_{max}=0.7$, $\alpha=13^\circ,$ and $\chi=35^{\circ}$. The observed J band profile, $Q_{\varphi}(r)/I_{\star}$, is plotted as a solid line for comparison. Units are given as relative flux per pixel of 3.6mas$\times$3.6mas}
    \label{fig:mod-profile}
\end{figure}

Figure~\ref{fig:mod-profile} shows the comparison between the observed J band profiles and the best disk grid model, with adjusted inner disk radius $r_i=0.38\arcsec$, which yields the rim radius of $r_w=0.58\arcsec$ for $\alpha=13^\circ$ and $\chi=35^\circ$. 
There is a good match between the observations and the model profiles at the inner edge, which is not resolved so that the profile shape up to the peak intensity is defined by the convolution. 
The observed disk shows the characteristics of a steep slope at $r_i$, which turns toward larger $r$ continuously into a flatter slope illuminated under grazing incidence. 
This cannot be reproduced by a single slope model that ends at a fixed radius, $r_w$; the agreement is less good farther out. The profile $Q_\varphi(r)$ of the best fitting model is with respect to the observations underestimating outside $r> 0.58\arcsec$, and overestimating the signal from about $r\approx 0.4\arcsec$ to $\approx 0.58\arcsec$. Of course, the disk-integrated polarized intensities calculated from the observed $Q_\varphi(r)/I_\star$ and fitted profiles $\tilde{Q}_\varphi(r)/I_\star$ (multiplied by $2\pi r$) are in good agreement, and one can expect that a more sophisticated model fitting would provide interesting information about the disk structure.

\subsection{Color dependence of disk radiation parameters}
To derive the dust properties in circumstellar disks, it is important to disentangle the parameter dependences related to the dust scattering from those related to the disk geometry for the measured quantities.  
Radiation parameters that depend mainly on the dust parameters are $\langle p_\varphi \rangle$ and $\Lambda_I$ as discussed in previous 
paragraphs. 
The wavelength dependence of the scattered light is another powerful tool to characterize the dust because we can assume that the scattering geometry is at least roughly the same for different colors \citep[see][]{Ma2022}. 

We derived a large difference for the polarized reflectivity between $\hat{Q}_\varphi/I_\star=0.92~\%$ for the R band and $1.51~\%$ for the J band in RX~J1604. 
This can be attributed mainly to the wavelength dependence of the dust scattering parameters. 
A reflectivity or polarized reflectivity that increases with wavelength has been reported previously for several protoplanetary disks \citep[e.g.,][]{Stolker2016, Monnier2019, Hunziker2021, Tschudi21, Mulders2013}.
A positive gradient for the reflectivity $(I_{\rm disk}/I_\star)_\lambda$ can be explained, if $\omega$ increases with $\lambda$, or if $g$ decreases with $\lambda$, or a combination of the two effects. 
Also, an increase in the scattering polarization $p_{\rm max}$ with wavelengths \citep[][Fig.~10]{Ma2022} could contribute significantly to the positive or red $(\hat{Q}_\varphi/I_\star)_\lambda$-gradient $\eta_{R/J}\approx 0.8$ (Eq.~\ref{Eq:refgradient}) for the polarized reflectivity of RX~J1604. 
It is interesting to note that a similar gradient of $\eta \approx 1$ was also derived of HD~142527 and HD~169142 \citep{Hunziker2021,Tschudi21}. There are more disk observations in the ESO archive suited for the measurements of the polarized reflectivity colors $\eta$ and an analysis of these data, which will put the obtained red color for RX J1604 into a more comprehensive context, is in preparation \citep{Ma_prep}.

Additional information on the dust in RX~J1604 follows from the $Q_\varphi/I_\star$ profiles, which show a clear correlation between the wavelength and the peak radius (Sect.~\ref{sect:rlambda}). 
This can be explained by the dust opacity, which decreases with $\lambda$ so that longer wavelength photons penetrate on average deeper into the surface before scatterings occur. 
This is the typical behavior for dust with a dominant population of small grains with a grain size distribution $n(a)\propto a^{-\gamma}$ and exponent $\gamma > 3$ like for interstellar dust \citep[e.g.,][]{Kruegel2003}. 

The frequently occurring dips in the brightness of the central star in RX~J1604 are attributed to the extinction by optically thin transients or passing clouds of hot dust near the star (Sect.~\ref{section:the-disk}).
These dips show also a stronger absorption at short wavelengths and also infer a predominance of small dust particles in the hot dust cloud \citep{Sicilia2020}. 


\section{Discussions}
\label{sect:discussions}
\subsection{Radiation parameters for the disk RX~J1604}
The RX~J1604 disk has a bright, well-defined inner wall with a large angular separation of $r\approx 0.4\arcsec$, which can be easily separated from the central star with high-resolution polarimetric imagers. 
The disk is seen pole-on and has a simple axisymmetric geometry, which allows an analysis of the disk reflectivity based on azimuthally averaged radial profiles following the study of \citet{Tschudi21}.
High-quality profiles can be measured within less than 10~minutes so that the impact of PSF variations caused by the atmospheric seeing and the variable AO performance can be accurately evaluated and corrected. 

Minor complications for RX~J1604 are the photometric variability of the central star because of transient dips and the shadows on the east and west side of the disk \citep{Ansdell2016, Pinilla2018}. Both effects are explained by unresolved hot dust near the star and the two effects are taken into account in our analysis.
We also investigate in detail the impact of polarimetric calibration errors and conclude that the apparent shadows in the R band data could be significantly affected by calibration errors.

We accurately measured the intrinsic radiation parameters $\hat{Q}_\varphi/I_\star$ for the R band and the J band, $I_{\rm disk}/I_\star$ for the J band, $\langle \hat{p}_\varphi \rangle$ for the J band and the H band, and radial profiles for the polarized intensity $\hat{Q}_\varphi(r)/I_\star$ (see Table~\ref{tab:fitting-parameters}). Accurate measurements means that the uncertainties for these parameters are only about $\pm 10~\%$ or even less and therefore very valuable for constraining strongly the dust in this disk.

In addition, we derive the ratio $\Lambda_I$ between the scattered radiation and the thermal emission of the dusty disk and therefore constrains strongly the single scattering albedo $\omega$ of the scattering dust. We also quantify accurately the wavelength dependence of the apparent location of the inner wall in scattered light. 
As already previously noticed by \cite{Pinilla2015}, the apparent radius is larger in the near-IR when compared to the visual because the dust extinction decreases with wavelengths. 
Our accurate radii $(r_{\rm peak}(\lambda))$ could be useful to describe the dust stratification in the inner wall of a transition disk.

This study on RX~J1604 indicates that an even higher measuring precision can be achieved with the same instrumentation. ZIMPOL observations taken under better seeing conditions and a more sophisticated strategy using reference star data from the same night should allow the extraction of the disk intensity $I_{\rm disk}/I_\star$ and the determination of $\langle p_\varphi \rangle$ for the R band. 
A better PSF calibration strategy for IRDIS should provide more accurate $Q_\varphi/I_{\rm disk}$ values. Taking data for different wavelengths during the same night or consecutive nights would also reduce the impact of the temporal variability of the shadows on the disk and improve the measurements of the spectral dependence of the reflected radiation. 
Thus, RX~J1604 is an ideal object for accurate determinations of the reflected light from circumstellar disks and the investigation of dust parameters for many wavelengths. The radiation parameters $\hat{Q}_\varphi/I_\star$, $I_{\rm disk}/I_\star$, $\langle \hat{p}_\varphi \rangle$, and the radial profiles $\hat{Q}_\varphi(r)/I_\star$ are the interface between the observations and the dust scattering models for the circumstellar disks and the results given in Table~\ref{tab:fitting-parameters} provide first benchmarks for RX~J1604, which should be complemented, checked, and improved with future measurements. 

\subsection{Comparison with existing dust models}
The properties of the scattering dust can be derived with a detailed quantitative comparison between model calculations and the disk radiation parameters derived from the observations. The disk of RX~J1604 is very favorable for such an analysis because it has a simple geometry seen pole-on that can be described well with axisymmetric disk models. In the ideal case the employed calculation should be based on physical dust models, for a disk geometry based on hydrodynamical simulations and state-of-the-art radiative transfer calculations. For RX~J1604 a comprehensive radiative transfer modeling for the thermal emission of the disk is presented in \citet[e.g.,][]{Woitke19}, but no results for the scattered radiation from the disk, nor the scattering properties of the modeled dust, are provided. 
For other disks there exist calculations that provide results for both, the thermal IR and the disk geometry as observed in scattered light \citep[e.g.,][]{Pinte2008, Monnier17, Monnier2019}. These studies consider the scattered light measurements for the intensity $I_{\rm disk}/I_\star$ or polarized intensity $\hat{Q}_\varphi/I_\star$, but they analyze their data with a model analysis focused on the hydrodynamical disk structure and not on the characterization of the scattering dust.

Rough comparisons between observations and simulations of the reflected light from the disk, which consider only the disk morphology or only the intensity of the disk cannot break the parameter ambiguities for the dust scattering, because many different dust models predict similar values for either $I_{\rm disk}/I_\star$ or $\hat{Q}_\varphi/I_\star$, or even both.
To break ambiguities one should compare as many disk parameters as possible in a quantitative way including an assessment of uncertainties for the measured values and of bias effects introduced by the model selection.

As a first approach, we compared the J band measurements from Sect.~\ref{section:discussion} with the disk models from \citet{Ma2022}, which are based on a three-parameter description of the dust scattering. This allows a rough exploration of a large parameter space from low to high albedo dust, with strong forward, quasi-isotropic, and low and high scattering polarization. 

The diagnostic power of the obtained J band measurements could be further increased by including the observed radial profile $Q_\varphi(r)/I_\star$ in comparison with simulations. Unfortunately, we lack systematic disk model calculations providing results for the radial profiles of the scattered light for different disk structures. 

We note that the parametric description of \citet{Ma2022} is a simplification that restricts the scattering phase functions for the intensity and polarization to a predefined one-parameter description. For example, this excludes dust that has not only a forward scattering peak from diffraction but also a back-scattering peak from reflection by the surface of highly structured large grains as seen for the zodiacal light or debris disks \citep[e.g.,][]{Ishiguro2013, Min2010, Chen2020}. 
In addition, this description allows unphysical parameter combinations such as strong forward scattering and diffraction (e.g., $g\approx 0.8)$ as expected for larger particles and a low albedo (e.g., $\omega\approx 0.2$).
This can be excluded for large particles because the contribution of the diffraction to the scattering albedo alone is at a level of $50~\%$. Therefore, there is probably always $\omega>0.5$ if the asymmetry is at a level of $g\approx 0.8$. 
Using physical dust scattering models is required to achieve significant progress in the characterization of dust in circumstellar disks. 

Mie scattering theory for spherical dust particles \citep[e.g.,][]{Monnier17, Bertrang2018} or scattering by hollow spheres \citep[e.g.,][]{Maaskant2013} are often used and these models provide good results for the modeling of the thermal emission from the dust in circumstellar disks. 
However, dust described by the Mie theory or hollow spheres usually do not explain the observations of the reflected intensity and polarization of protoplanetary disks, which often indicate for the near-IR (around $\lambda\approx 1-2\, \mu$m) the presence of dust with relatively high forward scattering $g\approx 0.5$ and high scattering polarization $p_{\rm max}\approx 0.5$.
This is not expected from Mie-scattering, because $g> 0.5$ is expected for large dust particles producing substantial forward diffraction while $p_{\rm max}\approx 0.5$ requires small spherical dust grains behaving like small Rayleigh-like scattering particles \citep[e.g.,][]{Min2016}. 

Therefore, dust aggregate models are proposed, which are dust particles with substantial small-scale structures \citep[e.g.,][]{Pinte2008}. 
Large aggregate particles can produce as a result of their size a strong forward scattering peak, but also a high scattering polarization, because the photon interactions with the small-scale structures behave similarly to the highly polarizing Rayleigh-scattering by small particles \citep[e.g.,][]{Kimura06, Min2016, Tazaki2016}. 
Many different types of dust aggregate models exist and it is not clear which one should be favored for the dust in protoplanetary disks \citep{Tazaki2019}. 
The dust scattering parameters derived for RX~J1604 in this work can therefore be very useful to narrow down the applicable dust models. 

Measurements of radiation parameters for the disk of RX~J1604 or other disks can be much more powerful for the derivation of the physical properties of the scattering dust if accurate data for different wavelengths are available and can be included in the analysis. For RX~J1604, we could derive a well-defined color dependence of $\eta \approx 0.8$ (Eq.~\ref{Eq:refgradient}) for the polarized reflectivity $\hat{Q}_\varphi/I_\star$, which indicates a red color, as observed in several other disks. According to the model simulations presented in \cite{Tazaki2019}, this is more compatible with lower porosity dust and less compatible with very porous dust.

For RX~J1604 also the wavelength dependence for the fractional polarization $\langle p_\varphi \rangle$ of the disk could be obtained using higher quality observations with SPHERE for the R band or the H band than for the data used in this work. 
In inclined disks, it is possible to measure the wavelength dependence of brightness differences between the front and back side, which are likely attributable to the wavelength dependence of the dust scattering asymmetry $g$. 
Such data would provide the wavelength dependence of the dust scattering parameters $\omega(\lambda)$, $g(\lambda),$ and $p_{\rm max}(\lambda)$ and strongly constrain physical dust models that should not fit the data at one wavelength only but at several wavelengths. 
For example, \cite{Tazaki2022} point to a possible systematic increase in $p_{\rm max}(\lambda)$ with the wavelength in the protoplanetary disk, which constrains the typical sizes of the monomers in the dust aggregates.

\section{Conclusions}
\label{sect:conclusions}
We present in this work quantitative measurements of the reflected radiation from the transition disk RX~J1604 taken from AO observations using VLT/SPHERE.\  Our main goal was to characterize the scattering dust in this disk.
We accurately measure the intrinsic radiation parameters $\hat{Q}_\varphi/I_\star=0.92\pm0.04\%$ for the R band and $1.51\pm0.11 \%$ for the J band, $I_{\rm disk}/I_\star = 3.9\pm0.5\%$ for the J band, $\langle \hat{p}_\varphi \rangle = 38\pm 4\%$ for the J band, and $42\pm2 \%$ for the H band. We derive the apparent disk albedo, combined with IR excess, as $\Lambda_I = 0.16\pm 0.08$. In addition, we derive a red color, $\eta \approx 0.8$, for the polarized reflectivity, $\hat{Q}_\varphi/I_\star$, from the R to the J band, which indicates the wavelength dependence of the dust parameters. 

We compared, as a first approach, the observational results with the grid of models from \cite{Ma2022}.\ They are simple parametric models that describe the disk geometry with an angular disk height, $\alpha$, and a wall slope, $\chi$ (the disk inclination is fixed to $i=7.5^\circ$) and describe the dust with a scattering albedo, $\omega$, an asymmetry, $g$, and a polarization, $p_{\rm max}$. Our work here shows that these simple disk models already provide very strong constraints on the dust scattering parameters for the J band measurements of the integrated disk intensity, $I_{\rm disk}/I_\star$, the polarized intensity, $Q_\varphi/I_\star$, and the apparent albedo, $\Lambda_I$. 

Particularly powerful for the dust characterization are the disk-averaged fractional polarization, $\langle p_\varphi \rangle$, and the apparent disk albedo, $\Lambda_I$, because these parameters depend mainly on the dust scattering and much less on the disk geometry, so parameter ambiguities are strongly reduced. With these models, the RX~J1604 measurements define tight relations between the dust scattering parameters $p_{\rm max}$ and $\omega$ from $\langle p_\varphi \rangle$ (Eq.~\ref{Eq:pmax}), and between $g$ and $\omega$ from $\Lambda_I$ (Eq.~\ref{Eq:LambdaI}). 
Using the same model grid, we also derived best fitting dust scattering parameters $\omega\approx 0.50$, $g\approx0.50$, and $p_{\rm max}\approx 0.70$ using an MCMC method. If the measuring uncertainties are taken into account, then all three values can move together up or down along a band in the 3D parameter space defined by $(\omega,g,p_{\rm max}),$ as described quite well by Eqs. 12 and 13. 

Quantitative photo-polarimetric analyses can and should be applied to different disk types, young and older disks, weakly and strongly illuminated disks, or disks with a lot of gas and disks depleted of gas. This could reveal systematic differences related to disk properties, and point to dust evolution effects that should also be understandable with the favored physical dust model. 

While many accurate measurements still need to be obtained, the prospects are good because the  instrumentation needed \citep[see the references in][]{Schmid2022} to obtain scattered light data from circumstellar disks like for RX~J1604 now exists. This  allows for detailed comparisons of high-quality measurements with sophisticated dust scattering models for several or even dozens of individual disks.

\begin{appendix}

\section{Polarimetric calibrations}
\label{App}
Quantitative polarimetry requires the assessment and the calibrations of the telescope and instrument polarization, and the corrections for the interstellar polarization and the intrinsic polarization. The following subsections describe these effects for the measurements of the integrated azimuthal polarization $Q_{\varphi}$ of RX~J1604 and discuss the possible impact on the azimuthal distribution $Q_\varphi(\varphi)$ of the disk polarization signal. 
It should be noted that both SPHERE subsystems, ZIMPOL and IRDIS, use the half-wave plate switches between $Q^+$ and $Q^-$ or $U^+$ and $U^-$ measurements for the compensation of the instrument-introduced polarization $p_{\rm inst}$. However, the concept for the correction for the telescope polarization is different for the two subsystems. 
\begin{figure*}
    \centering
    \begin{subfigure}{0.31\textwidth}
        \includegraphics[width=\textwidth]{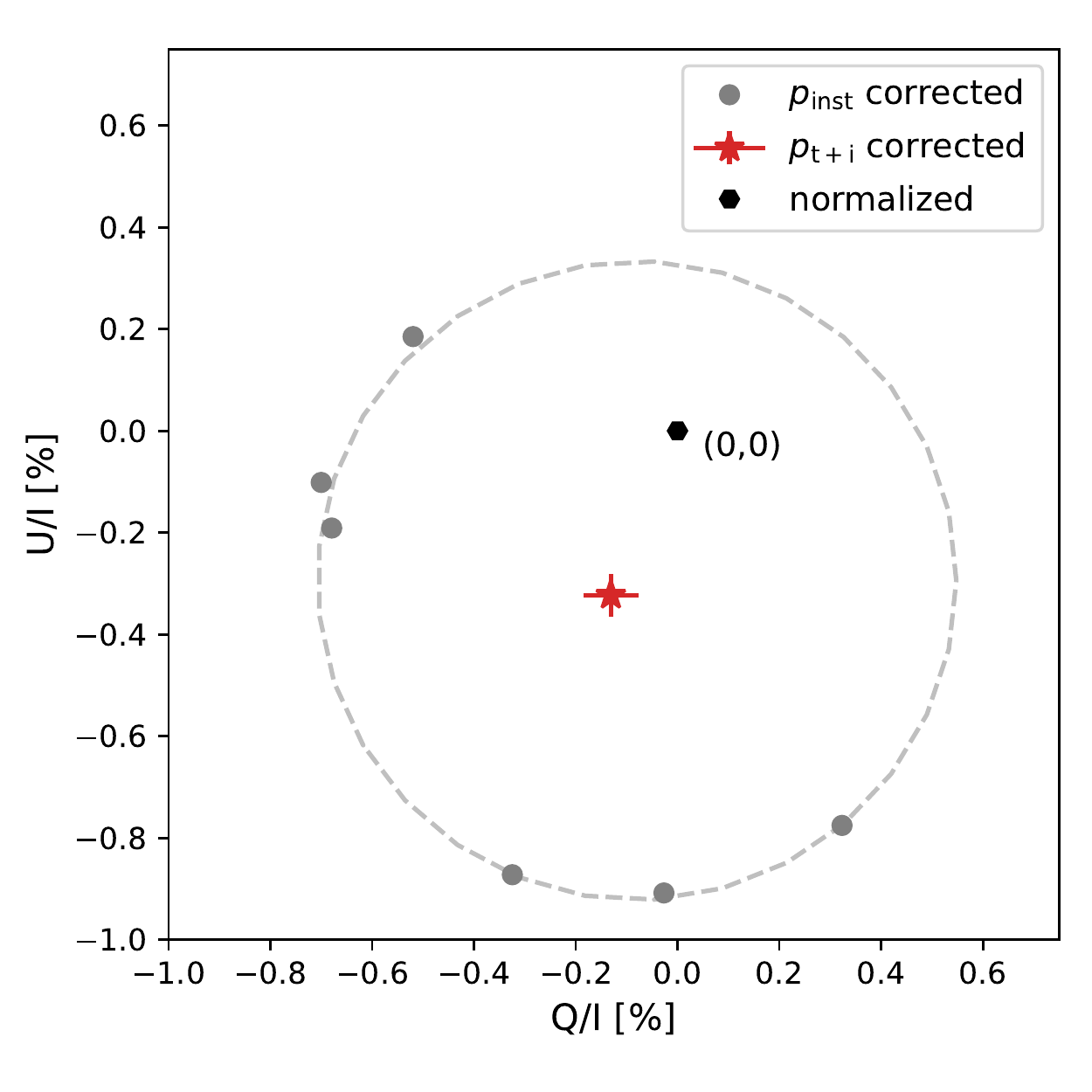}
    \end{subfigure}
    \begin{subfigure}{0.31\textwidth}
        \includegraphics[width=\textwidth]{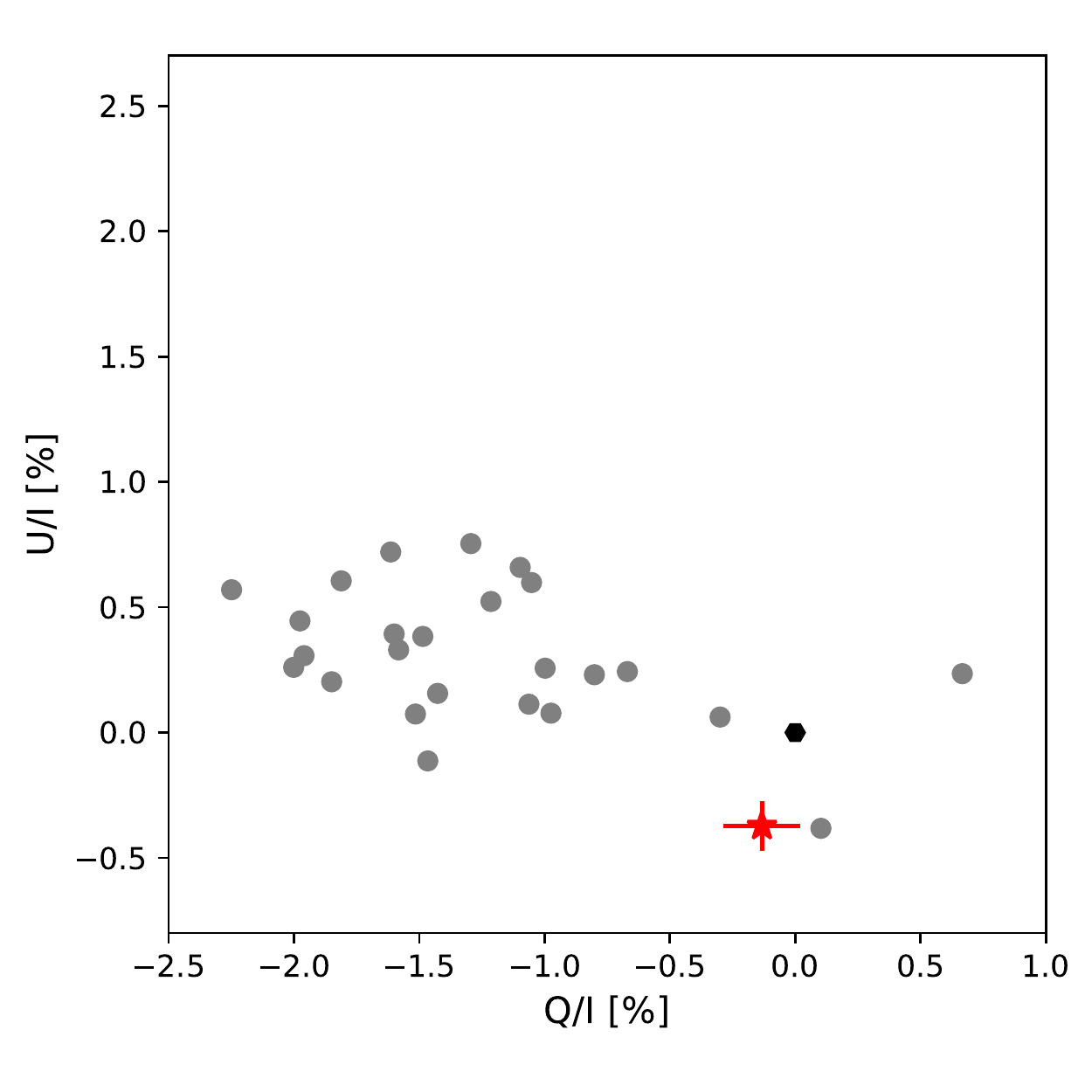}
    \end{subfigure}
    \begin{subfigure}{0.31\textwidth}
        \includegraphics[width=\textwidth]{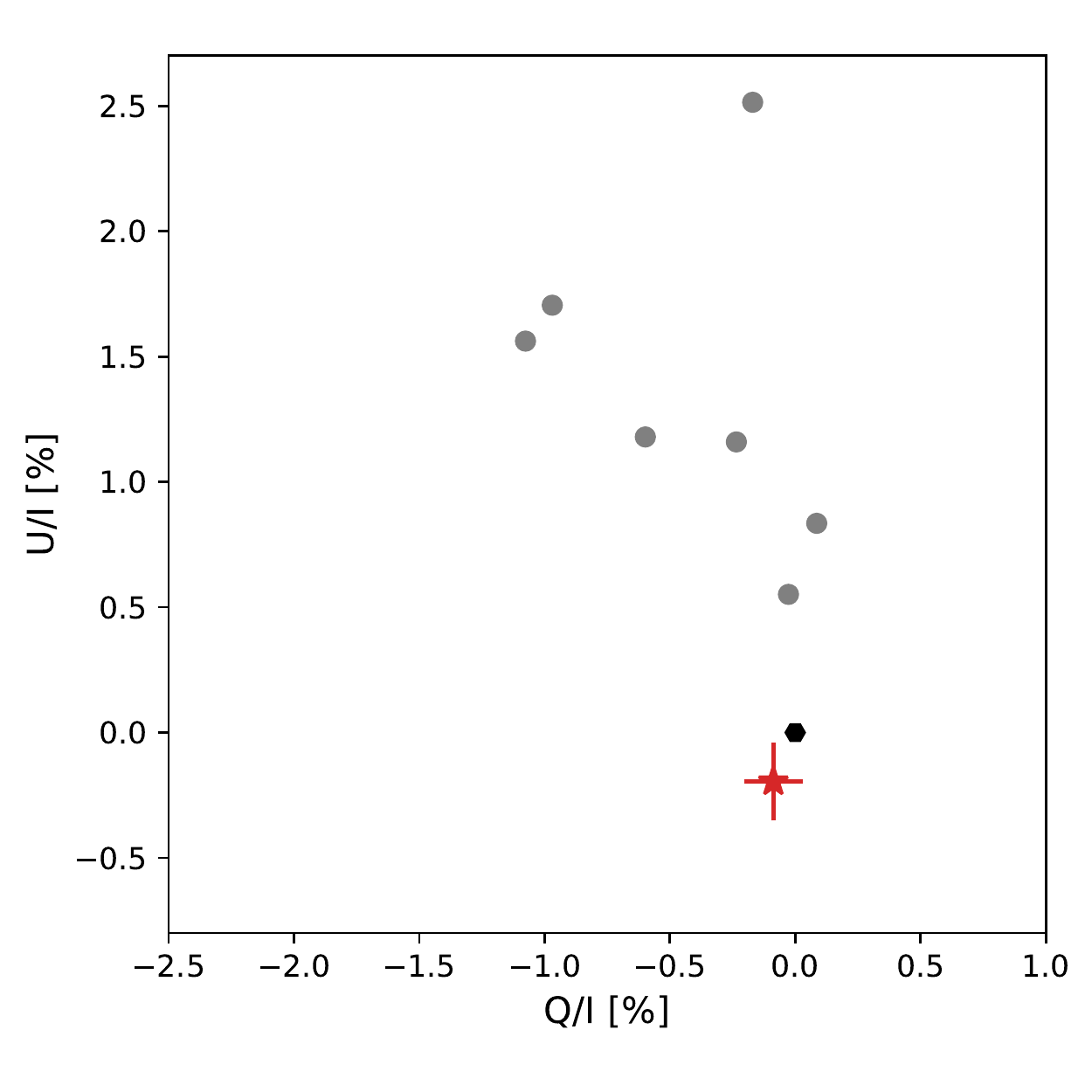}
    \end{subfigure}
    \caption{Polarization for the central star: measured values 
    (gray points), values corrected for the telescope and instrument polarization with error bars for the average values $(Q/I)_\star$, $(U/I)_\star$ (red points) for the ZIMPOL R band (left) and IRDIS J band (middle) and H band 
    observations (right).}
    \label{fig:correction}
\end{figure*}

\subsection{Calibration of the ZIMPOL data}
The instrumental polarization $Q_{\rm inst}$ and $U_{\rm inst}$ for ZIMPOL can be calibrated by the differences for polarimetric exposures $Q^+ - Q^-$ and $U^+ - U^-$, respectively, according to Eq.~19 in \citet{Schmid2018}. The telescope polarization $p_{\rm tel}$ is for the R band data almost constant, but the position angle of the telescope polarization $\theta_{\rm tel}(\theta_{\rm para})=\theta_{\rm para}+\delta_{\rm tel}$ rotates with the parallactic angle $\theta_{\rm para}$. 
The ZIMPOL observations of RX~J1604 were taken during the meridian passage, and because RX~J1604 passes close to the zenith the six polarimetric cycles cover a large range in parallactic angle from $\theta_{\rm para} = -106^\circ$ over $-180^\circ$ (=$+180^\circ$) to $108^\circ$. We measure the polarization of the central star for each cycle by selecting a small aperture of diameter $0.25\arcsec$ inside the disk and the measured polarizations are distributed on a half-circle in the $Q/I-U/I$-plane with a radius of $p_{\rm tel}$. We considered for this analysis the effect  of the strongly changing telescope orientation during an individual polarimetric cycle as described in the Appendix of \citet{Tschudi21}. 

Figure~\ref{fig:correction}{a} illustrates the best fitting circle for the telescope polarization model, where the origin of the circle defines 
the fractional polarization of RX~J1604 $(Q/I)_\star= -0.13 \pm 0.05~\%$ and $(U/I)_\star = -0.32\pm 0.04~\%$ or $p_\star =  0.35\pm 0.05 \%$ and $\theta_\star = 124 \pm 4 ^\circ$ with uncertainties derived from the distribution of the corrected values for the stellar polarization. The derived parameters for the telescope polarization $p_{\rm tel} =0.63 \%$, $\delta_{\rm tel} = 14.4^\circ$ are in good agreement with the values $0.55\%$, $12.6^\circ$ reported previously for the narrow band R filter \citep{Schmid2018}. 

\subsection{Calibration of the IRDIS data with IRDAP}
We used for the polarimetric calibration of the IRDIS data the IRDAP pipeline, which corrects the data according to a detailed Mueller Matrix model of the telescope and the instrument \citep{vanHolstein2020}. We considered that the IRDAP default aperture includes part of the RX~J1604 disk and hence the $p_\star$ measurements could be contaminated by the disk signal. 
Therefore, we used annular apertures with $0.12\arcsec<r<0.25\arcsec$ and $0.73\arcsec<r<1.5\arcsec$ where the stellar light dominates. 
The regions avoid the center, which could include systematic effects introduced by the coronagraphic mask, and the region where the polarization signal from the disk contaminates the stellar polarization.   

Figure~\ref{fig:correction}(b) and (c) show the distribution of measured or "raw" data points $(Q/I)$ and $(U/I)$ for the individual cycles, and the averages of the corrected values for individual observing nights with standard deviation $\sigma$ indicated as error bars.
The instrumental and telescope polarization effects for IRDIS are quite large and cannot be described with a simple circle in the $Q/I-U/I$-plane \citep{deBoer2020,vanHolstein2020}, but the good agreement for the resulting stellar polarization for the 5 epochs taken in the J band indicate that the telescope and instrument calibration procedure is reliable and provides good calibrations for IRDIS polarimetry. \\

The final results for the calibrated polarization for the central star of RX~J1604 are $p_{\star}(J)=0.42 \pm 0.12 \%$ with $\theta_{\star}(J)=126\pm10 ^\circ$ for the J band and $p_{\star}(H)=0.28 \pm 0.08 \%$ and $\theta_{\star}(H) =112 \pm 31^\circ$ for the H band. These values correspond to the red crosses in the $Q/I-U/I$-plane plots given in Fig.~\ref{fig:correction}(b) and (c).

As an additional test, we evaluated the polarization of the PSF reference star HD~150193 A and its companion B, because this system is a well-known high-polarization object. We measured for the A component $p_\star = 3.37 \pm 0.12 \%$ and $\theta_\star = 55.5 \pm 0.2^\circ$ and the B component
$p_\star = 3.13 \pm 0.21 \%$ and $\theta_\star = 56.2 \pm 0.5^\circ$ for the J band. 
This is consistent with the previous J band values $p_\star = 3.27\pm 0.06\%$, $\theta_\star = 57 \pm 1^\circ$ from \citet{Whittet92} and $p_\star = 3.14\pm0.02 \%$, $57 \pm 4^\circ$ compiled in \citet{Hales06}, both measuring the whole HD150193 A-B system with aperture polarimetry. 

\subsection{Interstellar polarization and normalization}
We found no previous calibrated polarization measurements for RX~J1604 in the literature. However, our measurements in the three bands, R, J, and H, characterize the stellar polarization of RX~J1604 well. 
We find no significant rotation in the position angle of polarization with wavelengths $\theta_{\star}(R)=124\pm 4^\circ$, $\theta_{\star}(J)=126\pm 10 ^\circ$ and $\theta_{\star}(H)=112\pm 31^\circ$ and the derived fractional polarizations $p_\star(R) = 0.35~\%$, $p_\star(J) = 0.42~\%$ and $p_\star(H) = 0.28~\%$ are both compatible with the typical wavelength dependence expected for interstellar polarization \citep[e.g.,][]{Serkowski75,Whittet92}. 
This interpretation is strongly supported by measurements of the polarization of stars within a few degrees and roughly at a distance $d=100$ to 150~pc, which show a similar $p_{\star}$ and $\theta_{\star}$ as RX~J1604. 
For example, HD~143275 at 150~pc has $p_{\star}=0.33\%$ with $\theta_{\star}= 121^\circ$ and HD~146029 at 183~pc has $p_{\star}=0.28\%$ with $\theta_{\star}= 110^\circ$ compiled in \cite{Heiles1999}, very much like RX~J1604. 

We conclude that the stellar polarization measured for RX~J1604 is predominantly caused by interstellar polarization. A contribution of a strong intrinsic polarization component ($>0.2~\%$) for the unresolved central source, which could be produced by dust scattering very close to the star, is according to this study very unlikely.

We first assumed that the polarization $p_\star$, $\theta_\star$ is only caused by interstellar polarization that adds just a fractional polarization offset $(Q/I)_{\rm is}$ and $(U/I)_{\rm is}$, such as the telescope and instrument polarization, $(Q/I)_{\rm t+i}$ and $(U/I)_{\rm t+i}$. 
Therefore, we can correct both these effects together by a normalization of the "raw" polarization of the central star, which can be achieved by scaling the linear polarization $I_0$ with respect to $I_{90}$ and $I_{45}$ with respect to $I_{135}$ so that $I_0^{\rm cor}=I_{90}^{\rm cor}$ or $Q^{\rm cor}=0$ and $I_{45}^{\rm cor}=I_{135}^{\rm cor}$ or $U^{\rm cor}=0$ as indicated by the black zero points in the panels of Fig.~\ref{fig:correction}. 
All the values for $Q_{\varphi}$ given in the Tables~\ref{tab:integrated-intensity-summary} are determined after normalization and are therefore based on the assumption that the central, unresolved stellar source has no intrinsic polarization. 

If the central, unresolved object in RX~J1604 has an intrinsic polarization (e.g., at a level of $p \approx 0.2~\%$) then the normalization would introduce an erroneous offset in the fractional polarization of the calibrated $(Q/I)$ and $(U/I)$ image. 
However, this has only a small impact on the disk-integrated azimuthal polarization $Q_{\varphi}$ for the disk like RX~J1604, which departs not much from an axisymmetric geometry. 
For example, only correcting the R band data of RX~J1604 for the telescope and instrument polarization $(Q/I)_{\rm t+i}$ and $(U/I)_{\rm t+i}$, but not accounting for the interstellar polarization, gives $Q_{\varphi} = 0.432\%$. 
This is practically identical to the derived value $0.439~\%$ obtained from the normalized polarization data. The difference is small because adding a contribution $Q/I$ to the $Q$-image will enhance on the left and right side of the star the positive signal $Q$ and therefore also the absolute value $|Q|$. 
Above and below the star the $Q$ signal is negative and a positive $Q/I$ contribution lowers the absolute $|Q|$-signal. For the integrated azimuthal polarization $Q_\varphi$ of an axisymmetric disk, these two effects compensate each other \citep[see][]{Hunziker2021}. 

\begin{figure}
    \centering
    \begin{subfigure}[!ht]{0.49\textwidth}
        \centering
        \includegraphics[width=0.95\textwidth]{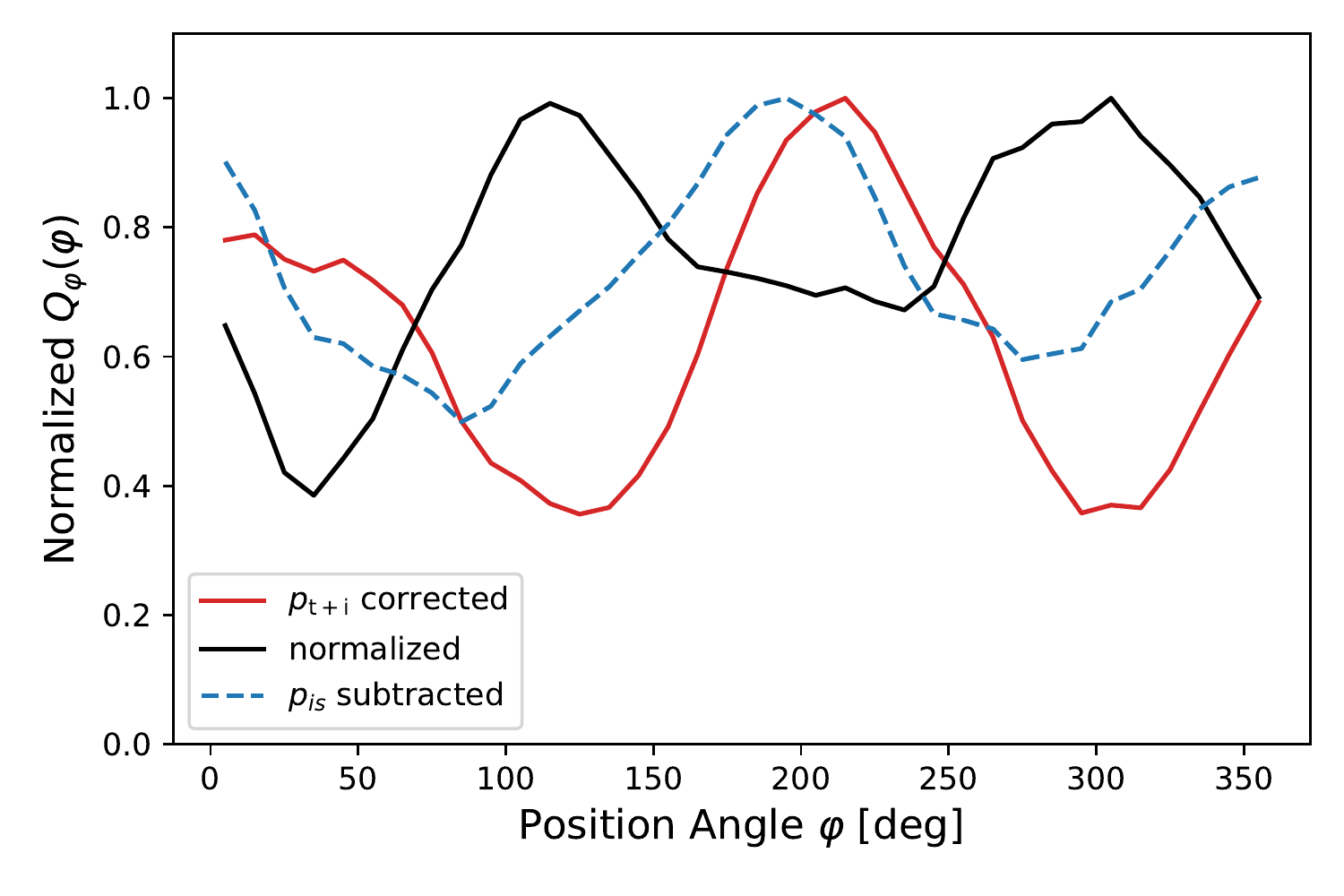}
    \end{subfigure}
    \caption{
    Azimuthal distribution of the R band disk polarization, $Q_{\varphi}(\varphi),$ of RX~J1604 for normalized data, which is equivalent to a correction of an interstellar polarization of $p_{\star}(R) = 0.35~\%$ and $\theta_{\star}(R)=124^\circ$ (black), and for data corrected for an interstellar polarization of $(Q/I)_{\rm is}(R) = 0.03~\%$ and $(U/I)_{\rm is}(R) = -0.22~\%$ to roughly match the shadow position in the near IR (blue). In both cases, the polarization was also corrected for the telescope polarization, $p_{t+i}$, which on its own would produce the profile $Q_{\varphi}^{t+i}(\varphi)$ shown in red.}
    \label{fig:azimuthal-profiles}
\end{figure}

\subsection{Normalization and azimuthal distribution of $Q_\varphi$}
\label{AppNormalization}
It should be considered that an erroneous polarimetric correction can have potentially a very significant effect on the azimuthal distribution of $Q_\varphi(\varphi)$ of the disk polarization. An offset $Q/I$ or $U/I$ can enhance the disk polarization on opposite sides of the star and reduce it in perpendicular directions. This effect can mimic shadows as observed in RX~J1604. 

We suspect that the positions of the shadows seen in the R band $Q_\varphi$-disk image from June 2015 in Fig.~\ref{fig:qphi-itot-image} (the same as presented in \citet{Pinilla2015} and \citet{Pinilla2018}) and also illustrated in the azimuthal profile $Q_\varphi(\varphi)$ in Fig.~\ref{fig:azimuthal-variance} are affected by such a calibration error. All these results were obtained after a polarimetric normalization.
If there is an intrinsic polarization for the central object, then the normalization introduces a calibration error. 

The problem for the R band data is illustrated in Fig.~\ref{fig:azimuthal-profiles}, which compares the azimuthal polarization profiles for the normalized data $Q^{\rm norm}_\varphi(\varphi)$ with the profile $Q^{\rm t+i}_\varphi(\varphi),$ which only corrects for the telescope and instrument polarization. The positions of the two minima in the R band data shift from about $\varphi\approx 35^\circ$ and $\approx 207^\circ$ for $Q^{\rm norm}_\varphi(\varphi)$ to about $\varphi\approx 120^\circ$ and $\approx 290^\circ$ for $Q^{\rm t+i}_\varphi(\varphi)$. 

It is clear from the study of \citet{Pinilla2018} that the derived positions of the shadows in the ZIMPOL R band image are not aligned with all the J band epochs presented in that work. Indeed, one can produce shadows at a position of roughly $\varphi\approx 90^\circ$ and $270^\circ$, like in all other RX~J1604 observations, by adopting for the central star in the R band an interstellar polarization of $(Q/I)_{\rm is} = 0.03~\%$ and $(U/I)_{\rm is} = -0.22~\%$ and an intrinsic polarization of $(Q/I)_\star = 0.10~\%$ and $(U/I)_\star = -0.10~\%$ as shown in Fig.~\ref{fig:azimuthal-profiles}.

It is well possible that such a small and hard-to-recognize intrinsic polarization of the central star is present. Therefore, the polarimetric normalization could change significantly the $Q_\varphi(\varphi)$ profile. This effect is expected to be particularly strong in the R band but much weaker in the near-IR data because (i) the intrinsic disk signal $Q_\varphi(\varphi)$ is much lower in the R band, (ii) the signal reduction by the PSF-smearing and polarimetric cancellation is much stronger (see the radial profiles in Fig.~\ref{fig:rad-model}) in the R band, and (iii) the signal of the "not perfectly" corrected polarization of the central star is particularly strong in the R band because the RX~J1604 disk coincides at this wavelength with the strong speckle ring bump of the 
stellar PSF. 
Therefore, one should consider the derived $Q_\varphi(\varphi)$ profile for the R band in Sect~\ref{section:azimuthal_variation} very cautiously. 
The corresponding effect for the IR bands is much lower and the impact is much less harmful and can be neglected for a first approximation for the derivation of the $Q_\varphi(\varphi)$ profiles for the J band and H band data.

In any case, these possible calibration errors do not affect significantly the disk-averaged results in Table~\ref{tab:fitting-parameters}, even for the R band data, because the net effect for the disk-integrated polarization $\hat{Q}_\varphi/I_\star$ can be neglected for axisymmetric scattering geometries like for RX~J1604 as discussed above.

\begin{acknowledgements}
We sincerely appreciate the valuable comments of the anonymous referee on the improvement of our manuscript. This work has been carried out within the framework of the NCCR PlanetS supported by the Swiss National Science Foundation under grants 51NF40\_182901 and 51NF40\_205606. CT and HMS acknowledge the financial support by the Swiss National Science Foundation through grant 200020\_162630/1. This work is based on data obtained from the ESO Science Archive Facility. 
\end{acknowledgements}

\end{appendix}

\bibliographystyle{aa} 
\bibliography{biblio}      
\end{document}